\documentclass[fleqn,usenatbib]{mnras}

\usepackage{newtxtext,newtxmath}

\usepackage[T1]{fontenc}

\DeclareRobustCommand{\VAN}[3]{#2}
\let\VANthebibliography\thebibliography
\def\thebibliography{\DeclareRobustCommand{\VAN}[3]{##3}\VANthebibliography}

\usepackage{graphicx}	
\usepackage{amsmath}	
\usepackage{orcidlink}

\def\farcs{\hbox{$.\!\!^{\prime\prime}$}}

\title[Reverberation Mapping of a Changing-State Quasar]{Probing the origin of the two-component structure of broad line region by reverberation mapping of an extremely variable quasar}

\author[S. Nagoshi et al.]{
Shumpei Nagoshi\orcidlink{0000-0003-2650-4322},$^{1}$\thanks{E-mail: shumpei@kusastro.kyoto-u.ac.jp}
Fumihide Iwamuro\orcidlink{0000-0002-7419-2629},$^{1}$
Satoshi Yamada\orcidlink{0000-0002-9754-3081},$^{2}$
Yoshihiro Ueda\orcidlink{0000-0001-7821-6715},$^{1}$
\newauthor
Yuto Oikawa,$^{1}$
Masaaki Otsuka\orcidlink{0000-0001-7076-0310},$^{1,3}$
Keisuke Isogai\orcidlink{0000-0002-6480-3799}$^{1,3,4}$,
Shin Mineshige$^{1}$
\\
$^{1}$Department of Astronomy, Kyoto University, Kyoto 606-8502, Japan\\
$^{2}$RIKEN Cluster for Pioneering Research, 2-1 Hirosawa, Wako, Saitama 351-0198, Japan\\
$^{3}$Okayama Observatory, Kyoto University, 3037-5 Honjo, Kamogata-cho, Asakuchi, Okayama 719-0232, Japan\\
$^{4}$Department of Multi-Disciplinary Sciences, Graduate School of Arts and Sciences, The University of Tokyo, 3-8-1 Komaba, Meguro, Tokyo 153-8902, Japan\\
}

\date{Accepted XXX. Received YYY; in original form ZZZ}

\pubyear{2022}

\begin{document}
\label{firstpage}
\pagerange{\pageref{firstpage}--\pageref{lastpage}}
\maketitle

\begin{abstract}

The physical origins of quasar components, such as the broad line region (BLR) and dust torus, remain under debate. To gain insights into them, we focused on Changing-State Quasars (CSQs) which provide a unique perspective through structural changes associated with accretion disk state transitions. We targeted SDSS J125809.31+351943.0, an extremely variable CSQ, to study its central core structure and kinematics. We conducted reverberation mapping with optical spectroscopy to explore the structure of the BLR and estimate the black hole mass. The results from H$\beta$ reverberation mapping indicated a black hole mass of $10^{9.64^{+0.11}_{-0.20}}\rm{M_\odot}$. Additionally, we analyzed variations in the optical to X-ray spectral indices, $\alpha_{\rm{ox}}$, before and after the state transition, to investigate the accretion disk. These variations in $\alpha_{\rm{ox}}$ and the Eddington ratio (from 0.4 \% to 2.4 \%) exhibitied behavior similar to state transitions observed in X-ray binary systems. Spectral analysis of H$\beta$ revealed a predominantly double-peaked profile during dim periods, transitioning to include a single-peaked component as the quasar brightened, suggesting that H$\beta$ contains a mixture of two components. Each of these components has its distinct characteristics: the first is a double-peaked profile that remains stable despite changes in the accretion rate, while the second is a variable single-peaked profile. Using time lags from reverberation mapping, we estimated the spatial relationships between these BLR components, the accretion disk, and the dust torus. Our results suggest that the BLR consists of two distinct components, each differing in location and origin.

\end{abstract}






\section{Introduction} \label{sec:Introduction}


The broad line region (BLR) is one of the key components to solve several significant problems in astronomy. First, knowing the mechanism of the black hole's gaining mass will lead to understanding the formation process of supermassive black holes and the co-evolution with the host galaxy \citep{1998AJ....115.2285M}. The information on the structure and kinematics of the BLR is essential to investigate the black hole evolution because it is generally assumed when astronomers estimate the black hole mass of distant quasars. Secondly, the precise picture of quasars is also valuable for measuring the distances from our galaxy. A controversial issue in observational cosmology is the Hubble tension problem \citep[see][as a recent review]{2022NewAR..9501659P}, where the measured Hubble constant ($H_0$) with cosmic microwave background \citep{2020A&A...641A...6P} and with the distance ladder \citep[e.g.,][]{2022ApJ...934L...7R} differ by more than $5\sigma$. In this context, distance measurements using the BLR of quasars \citep[e.g.,][]{2020NatAs...4..517W} are expected to be a clue to solving the problem, as they can estimate $H_0$ independently of traditional methods. For these reasons, we believe it is important to know about the structure of the BLR.

Broad emission lines (BEL) are major characteristics of Type~1 quasars. A picture based on the unified model \citep{1993ARA&A..31..473A, 1995PASP..107..803U}, in which dust is distributed in a torus structure around the accretion disk and the BLR, is widely accepted as the general structure of quasars. This simple picture can explain diverse spectra of quasars, Type~I and II, only by one parameter of viewing angle. The detailed structures, however,  are still under discussion as well as the motion and origin of the BLR \citep[see][as a review]{2009NewAR..53..140G}.

Reverberation mapping (RM) is one of the most promising methods for studying the quasar structure \citep[see][and references therein]{2021iSci...24j2557C}. Quasars randomly change luminosity at all wavelengths in diverse timescales and amplitudes.
Using that feature, RM is a method to estimate the size of the BLR by measuring the response time between the variation in the continuum emitted from the accretion disk and the variation in the BEL emitted from the BLR. The virial black hole mass can be calculated by the measured size of the BLR and its velocity, which is thought to be the most precise method for measuring the black hole mass of distant quasars. An important result of the RM is the discovery of a strong correlation between luminosity and the BLR radius \citep[L-R relation;][]{1991ApJ...370L..61K, 2000ApJ...533..631K, 2005ApJ...629...61K}, which has been widely used to estimate black hole mass from a single spectrum \citep[e.g.,][]{2011ApJS..194...45S}. RM can provide more detailed information by increasing the velocity resolution and observation frequency. As usual, emission lines are often divided into several velocity components to estimate the qualitative structure and kinematic information of the BLR \citep[velocity-resolved RM;][]{2016ApJ...820...27D, 2018ApJ...866..133D}. By using the information derived from each component, it will be possible to elucidate the structure and kinematics of the BLR. For example, if the red wing of the BEL lags behind the blue one, we can infer that outflow motion is dominant in the BLR. To investigate the properties of the BLR in detail, we adopted RM in this study.

Understanding the properties of the BLR is necessary for accurate estimation of the black hole mass \citep{1999ApJ...521L..95P, 2000ApJ...540L..13P}, which is a fundamental physical quantity of a quasar, because measurement of the black hole mass by RM depends on the structure, mainly the viewing angle, of the BLR. In RM, the black hole mass is calculated with the following equation;
\begin{equation}
\label{equation:BHmass}
 M_{\rm{BH}} = f \frac{c \tau \Delta V^2}{G},
\end{equation}
where $c$ is the speed of light, $G$ is the gravitational constant, and $\Delta V$ is the velocity of the BLR gas. The $f$ is a factor that reflects the structure, kinematics, and inclination of the BLR. 
The problem is that the $f$ factor is supposed to take different values for each object, because at least inclination is different from object to object, while it is currently assumed as the same value. Furthermore, in many cases, the radius of the BLR is estimated from a single spectroscopic observation using the \textit{empirical} L-R relation.
We express concern that discussions premised on such a series of assumptions may potentially lead to misleading interpretations. To solve these problems, it is necessary to create models that are closer to reality and check them against observations, but there are many observational uncertainties. Therefore, the purpose of this study is to gain insight into the structure of the BLR.

To understand the structure and kinematics of the BLR in detail, consideration of the origin of the ionized gas in the BLR is important. There are two major hypotheses for the origin of the BLR gas. One is from the accretion disks as the disk-wind \citep[e.g.,][]{2011A&A...525L...8C} and the other is the inflow (including tidally disrupted clumps) from the dust torus \citep[e.g.,][]{2009NewAR..53..140G}. On one hand, if the gas is supplied by the disk wind from the accretion disk, it is expected that the amount of disk wind changes with the mass accretion rate of the accretion disk, resulting in a variation in the BEL intensity \citep[][]{2014MNRAS.438.3340E}. On the other hand, some studies have suggested that tidally disrupted dust torus gas is responsible for the asymmetric component of the BEL \citep[][]{2017NatAs...1..775W}. As the observational results are different from object to object, the origin of the BLR is still under discussion. For example, the velocity resolved RM results revealed that some objects exhibited outflow features \citep[e.g.,][]{2019ApJ...887..135L, 2021ApJ...918...50L} while some exhibited inflow features \citep[e.g.,][]{1988ApJ...325..114G, 1989ApJ...345..637K, 2010ApJ...721..715D, 2013ApJ...764...47G, 2016ApJ...820...27D, 2019ApJ...876...49Z, 2021ApJ...918...50L, 2022ApJS..262...14B}, or both \citep[e.g.,][]{2018ApJ...869..142D, 2020ApJ...905...77B, 2021ApJ...918...50L}. A possible scenario is that multiple origins may be mixed. The key is to sort out what could exist as an origin and to what extent it is dominant.




We suggested in \cite{2022PASJ...74.1198N} that an ideal target for investigating the origin of the BLR is Changing-State (Look) Quasars\footnote{Strictly speaking, the definitions of Changing-Look Quasars and Changing-State Quasars are different \citep[see a review of][]{2022arXiv221105132R}. However, for the purpose of this paper, they are treated similarly.} (CSQs). We focus on the variability of BEL and continuum luminosity, which show significant variations ($>30\%$) in the BELs and in mid-infrared luminosity \citep{2015ApJ...800..144L, 2016MNRAS.457..389M, 2019ApJ...874....8M, 2016ApJ...826..188R, 2017ApJ...835..144G, 2018ApJ...858...49W, 2018MNRAS.480.3898N, 2020MNRAS.498.2339R, 2020MNRAS.491.4925G, 2021PASJ...73..122N, 2021PASJ...73..596W, 2022ApJ...933..180G}. This significant variation represents the structural transformation process, which is an ideal feature for investigating each component of the unified model. In particular, if we can observe the emergence process of the BLR, we can place restrictions on the origin of the BLR gas. Although the variation process of accretion disk also provides essential information, the observational studies are still not enough for quasars because of their long variation timescale. Therefore, the purpose of this study is to apply RM to one of the CSQs \citep{2019ApJ...885...44D, 2022ApJ...936...75L} together with X-ray observations to gain insight into the physical origin of the BLR and the physical process of drastic variation in the accretion disk.

In this paper, we study the BLR structure of SDSS J125809.31+351943.0 (hereafter J1258; $z = 0.30939$), which was discovered by \citet{2021PASJ...73..122N} as the object that experienced the most significant state transition in history (optical brightness variation of about 4 magnitudes in about 30 years). The significant brightness variation is expected to have a large signal-to-noise ratio of the varied component in the spectrum, making it a suitable target for RM. 

 Each section is organized as follows. Section~\ref{sec:Data} describes our observations and the archival data we used. Section~\ref{sec:Analysis_Result} describes the results of the RM analysis and the comparison of SED before and after the extreme variation. Section~\ref{sec:Discussion} discusses the black hole mass, the cause of the variation, and the quasar structure based on the above results. The cosmological parameters are assumed to be $ H_0 = 70$ $
\rm kms^{-1}Mpc^{-1}$, $\Omega_M = 0.3$ and $\Omega_{\Lambda} = 0.7$ based on the $\rm{\Lambda CDM}$ cosmology throughout this paper.  All errors that appear in this paper represent $1\sigma$ errors.

\section{Observations and Data Reduction} \label{sec:Data}
This section explains the data we used. Firstly, we describe the optical/mid-infrared photometry data, which mainly correspond to the continuum emission from accretion disk/dust torus, respectively. Secondly, the spectroscopic data mainly used to derive the light curve of H$\beta$ are explained. Finally, we describe the X-ray observations used to investigate the accretion state of the inner part of the accretion disk.

\subsection{Optical Photometry} \label{sec:Optphot}  
We included archival data of photometric observations, because the cadence of spectral observations in this study is insufficient to measure time-lag. The catalog data used were the Catalina Real-time Transient Survey (CRTS; \citealt{2009ApJ...696..870D}) and the All Sky Survey Automated Survey for Super-Nova (ASAS-SN; \citealt{2017PASP..129j4502K}). CRTS is unfiltered photometric data, while ASAS-SN observations mainly use $g$-band and $V$-band filters. Observations between different filters show consistent patterns of light curves, but a certain amount of offset exists. Therefore, an offset was applied to match the mean values in the overlapping period of the two light curves. Detailed analysis is described in \cite{2021PASJ...73..122N}. The essential information to measure the time-lag is the patterns of brightness variation. Therefore, we treat the combined light curve as a continuum light curve of J1258 in this analysis, though it contains values of different filters.

\subsection{Mid Infrared (WISE) photometry} \label{sec:Midirphot} 
The Wide-field Infrared Survey Explorer (WISE; \citealt{2010AJ....140.1868W}), which is a mid-infrared satellite launched in December 2009 and observed the entire sky in four bands (3.4, 4.6, 12, and 22 $\rm{\mu m}$), also observed J1258. WISE operations were temporarily terminated in February 2011, but observations resumed in December 2013 as the NEOWISE project \citep{2014ApJ...792...30M}. NEOWISE has two bands of 3.4 and 4.6 $\rm{\mu m}$ and observed the entire sky, visiting each area in the sky about ten times every six months. 

As the information of J1258, we referred to the single-exposure profile-fit magnitude within 3'' from RA = 12:58:09.31 and DEC = +35:19:43.03. From these values, we selected those with no contamination and confusion flags (``cc$\_$flags'' = 0000). In the present analysis, we used the combined data with each bin accumulated for six months period, sufficient for the variation timescale of our interest.

\subsection{Spectroscopy 1: Sloan Digital Sky Survey} \label{sec:Spec_SDSS}
J1258 has been spectroscopically observed twice by the SDSS, in April 2006 and April 2016. The SDSS \citep{2000AJ....120.1579Y} is based on a photometric survey using a 2.5-m wide-field telescope \citep{2006AJ....131.2332G} equipped with 30 $\rm{2k \times 2k}$ CCDs and following spectroscopic observations. Between the two observations, the spectroscopic instrument was upgraded, the wavelength range was extended and the thickness of the fiber on the SDSS plate was changed from 3$^{\prime\prime}$ to 2$^{\prime\prime}$ with the upgrade of the spectrograph.

\subsection{Spectroscopy 2: LAMOST} \label{sec:Spec_LAMOST}
J1258 was observed on April 2017, by the Large Sky Area Multi-Object Fiber Spectroscopic Telescope (LAMOST; \citealt{2012RAA....12.1197C, 2012RAA....12..723Z}), a Schmidt telescope with a primary mirror size of $6.67 \rm{m} \times 6.05 \rm{m}$. The wavelength ranges from 370 nm to 900 nm, and the wavelength resolution is about 1000. The data is reduced with LAMOST pipelines \citep{2015RAA....15.1095L}.

\subsection{Spectroscopy 3: MALLS} \label{sec:Spec_MALLS} 
We performed two spectroscopic observations in December 2018, and May 2019 with the slit spectrograph called ``MALLS'' (Medium And Low-dispersion Long-slit Spectrograph) installed in the 2-meter telescope in Nishi Harima Astronomical Observatory in Japan. The grating was 150 l/mm, giving a spectral resolution of $\sim 600$, with the 1\farcs{}2 slit width.
We executed five 1200-second exposures for two nights. The observed data were reduced with the standard processing of slit spectroscopy with IRAF (dark subtraction, flat correction, matching sky subtraction, wavelength correction, and flux correction using the spectrophotometric standard). 

\subsection{Spectroscopy 4: KOOLS-IFU} \label{sec:Spec_KOOLS} 
We performed spectroscopic monitoring observation in 2020 using the low-resolution spectroscopic data from KOOLS-IFU (Kyoto Okayama Optical Low-dispersion Spectrograph with optical-fiber Integral Field Unit; \citealt{2019PASJ...71..102M}) installed in the 3.8-m Seimei Telescope \citep{2020PASJ...72...48K} at Okayama Observatory of Kyoto University. The data are reduced with KOOLS-IFU pipeline\footnote{We used the pipeline here \url{http://www.kusastro.kyoto-u.ac.jp/~iwamuro/KOOLS/index.html}}.

\subsection{X-ray} \label{sec:xray} 
The target was observed twice using the ROSAT \citep{1982AdSpR...2d.241T} Position Sensitive Proportional Counter (PSPC) detector in 1991 December (ObsID = RP600164N00) and 1992 December (RP701178N00) with net exposures of 16.2~ks and 3.9~ks, respectively.
The ROSAT data were processed with Standard Analysis Software System (SASS).
The source spectra were extracted from a circular region with a radius of 150\arcsec\ centered on the source by using XSELECT.
The background spectra were selected from a partial annulus region with inner and outer radii of 
225\arcsec\ and 450\arcsec\ by excluding the region of a nearby X-ray source (1RXS J125740.2+351334; \citealt{1999A&A...349..389V}).
We used the standard response matrix (`pspcb\_gain2\_256.rsp') and generated the ancillary response files using \textsf{pcarf} command in HEASOFT v6.25.

We also used the data of Swift/X-Ray Telescope (XRT; \citealt{2005SSRv..120..165B}) in 2021 February (ObsID = 14079001) with a net exposure of 1.9~ks, covering the $\sim$0.3--10~keV band.
The data reduction was performed with the XRTPIPELINE v0.13.4.
The source spectrum was taken from a circular region of 45\arcsec\ radius, while the background was from an annulus between 90\arcsec and 150\arcsec\ radii.

\section{Data Analysis and Results} \label{sec:Analysis_Result}

In this section, we explain the analysis of optical spectra to investigate the variation of $\rm{H\beta}$. Then, we measured the time-lag of $\rm{H\beta}$, $W1$, and $W2$ to the continuum.

\subsection{Variation of \texorpdfstring{$\rm{H\beta}$}{H beta}} \label{sec:Variation_Hb}

To obtain accurate light curves, we calibrated the spectral flux assuming [OIII]5007 luminosity did not change during all observations. The region where [OIII]5007 is emitted extends over $\sim100$ pc, so it's often assumed that the intensity remains unchanged over a few decades of observations. We fitted the spectra to determine their [OIII]5007 flux with PyQSOfit \citep{2018ascl.soft09008G}. The components applied in the fitting are a power-law continuum, three Gaussian components for the broad $\rm H\beta$, and one Gaussian each for the narrow $\rm{H\beta}$, [OIII]4959/5007. First, the continuum is fitted with power-law using the window of 4450--4500~\r{A} and 5150--5250~\r{A} in the rest frame, and then the continuum component is subtracted from the spectrum. We use these continuum-subtracted spectra to fit the emission lines. The velocity offsets and the line width for [OIII]4959/5007, and the narrow $\rm{H\beta}$ are set to be equal, and the flux ratio was fixed to 1:3:0.07. Here, the ratio of the narrow component of $\rm{H\beta}$ was obtained by fitting the first SDSS spectrum, MJD 53848, with free parameters. We fixed the ratio of the narrow emission lines to this value in the fitting of the other spectra. After the fitting, we normalize all flux density using the flux of [OIII]5007 calculated with the Gaussian fitting. The total $\rm H\beta$ flux is measured by simply integrating the subtracted spectra in 4611--5111~\r{A} after the flux normalization. Each measurement error was estimated using the Monte Carlo method, which consists of 100 fitting calculations repeated with random noise based on the error of the spectrum. The value of the light curve of $\rm H\beta$ is summarized in the Appendix (Table~\ref{tab:obslog_table} and Figure~\ref{fig:lightcurves}). 

\begin{figure}
	\includegraphics[width=\columnwidth]{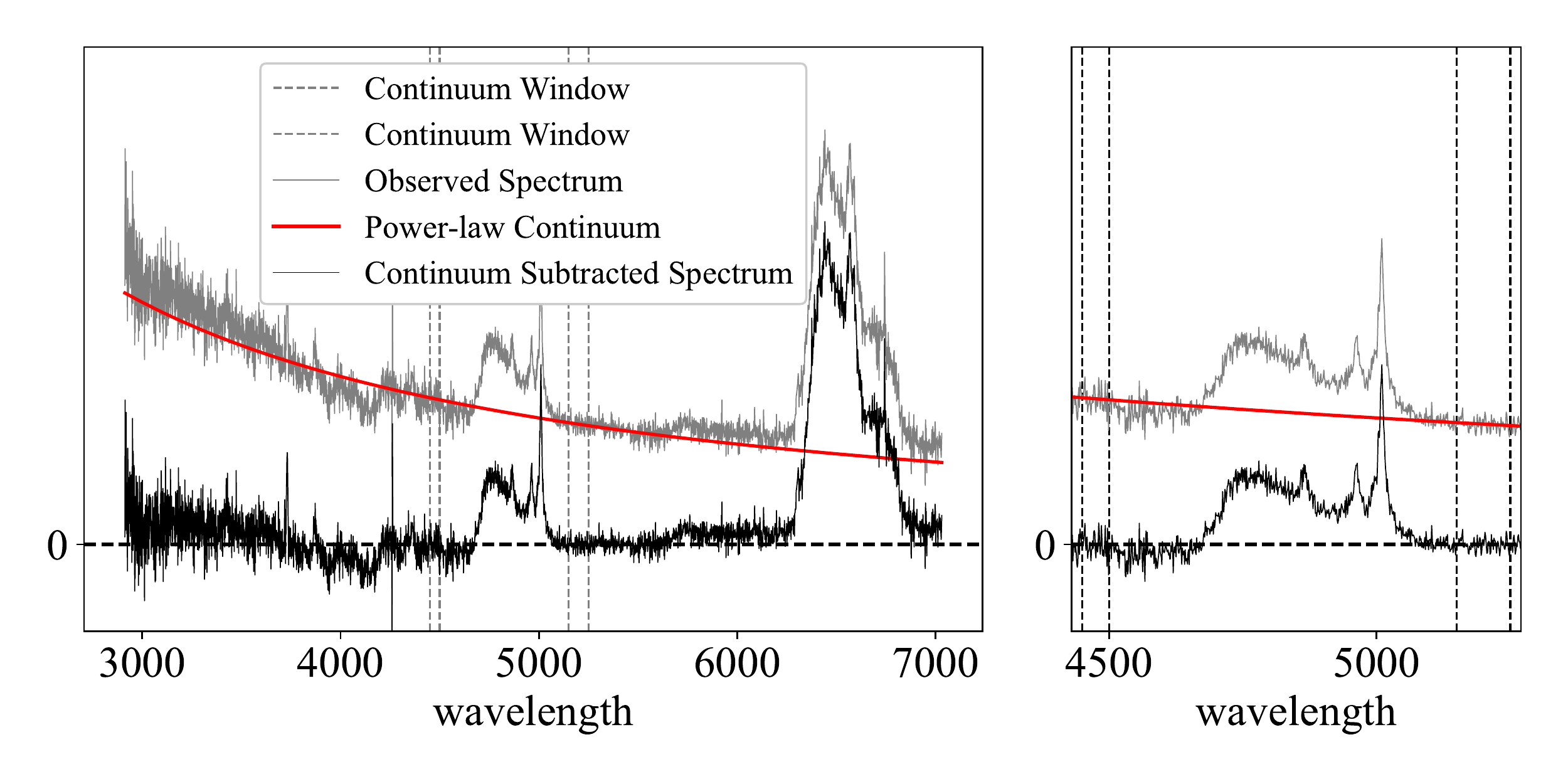}
    \caption{Example of power-law continuum subtraction. The original spectrum is shown in gray, and the fitted continuum component is represented in red. The black dashed lines indicate the wavelength windows used for the continuum fitting, while the black solid line shows the spectrum after subtracting the continuum component. The left panel displays the entire wavelength range, while the right panel focuses on the region around $\rm{H\beta}$.}
    \label{fig:fitting_example}
\end{figure}

\begin{figure}
	\includegraphics[width=\columnwidth]{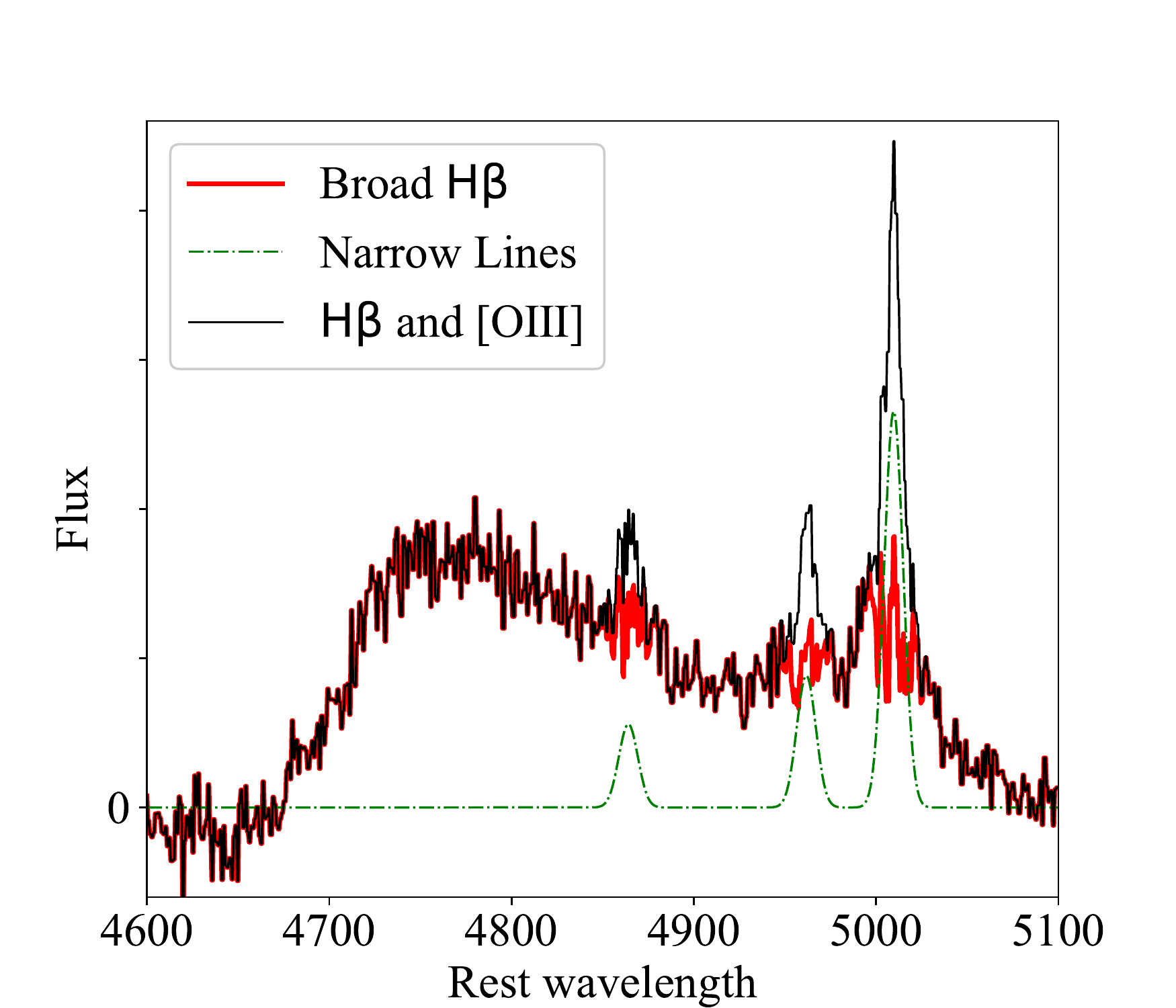}
    \caption{Example of emission line decomposition. The black line presents the emission line spectrum with the continuum subtracted. The red line represents the spectrum after subtracting the narrow emission line components, isolating the broad $\rm{H\beta}$. The spectrum of the narrow line components is illustrated with a green dashed line. We have assumed that the Narrow Line components remain constant throughout the observation period.}
    \label{fig:linefitting_example}
\end{figure}

\begin{figure}
	\includegraphics[width=\columnwidth]{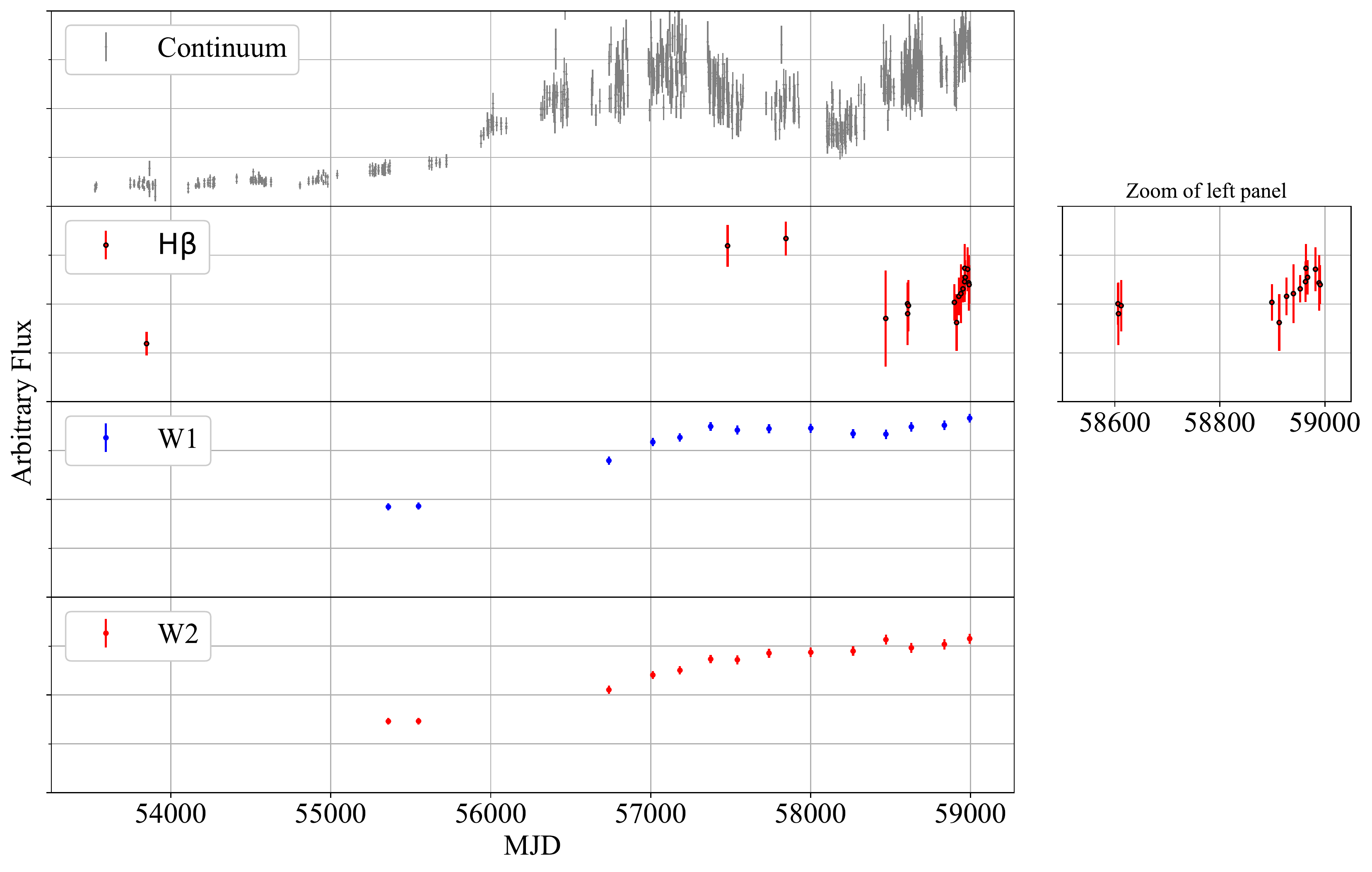}
    \caption{The light curves of continuum (top), $\rm{H\beta}$ (the second), WISE $W1$ (the third), and WISE $W2$ (bottom) of J1258.}
    \label{fig:lightcurves}
\end{figure}

We also calculated the mean spectrum and root mean square residual spectrum (RMS spectrum) to visualize the typical shape of the emission line and variable component \citep{2004ApJ...613..682P}. Because of the non-uniform time periods between observations (1 day to 3633 days), the calculation is weighted by the length of the observation period as a percentage of the light curve. The definition formula is as follows.
\begin{equation}
    \begin{split}
        f_{\rm{mean}} = \frac{1}{2(t_N - t_1)} \Bigl( f_1 (t_{2}-t_{1}) + f_N (t_{N}-t_{N-1})\\ + \sum_{i = 2}^{N-1} f_i (t_{i+1}-t_{i-1}) \Bigr)
    	\label{eq:meanspec}
    \end{split}
\end{equation}

\begin{equation}
    \begin{split}
        f_{\rm{RMS}} = \Biggl( \frac{1}{2(t_N - t_1)} \Bigl( (f_1-f_{\rm{mean}})^2 (t_{2}-t_{1})\\ + (f_N-f_{\rm{mean}})^2 (t_{N}-t_{N-1}) \\ +  \sum_{i = 2}^{N-1} (f_i-f_{\rm{mean}})^2 (t_{i+1}-t_{i-1}) \Bigr) \Biggr)^{\frac{1}{2}}
    	\label{eq:RMSspec}
    \end{split}
\end{equation}

Here, $f$ represents each spectrum, $t$ represents the date of each observation, and the numbers in the subscripts indicate the order of each spectrum by date of observation. 

\begin{figure}
	\includegraphics[width=\columnwidth]{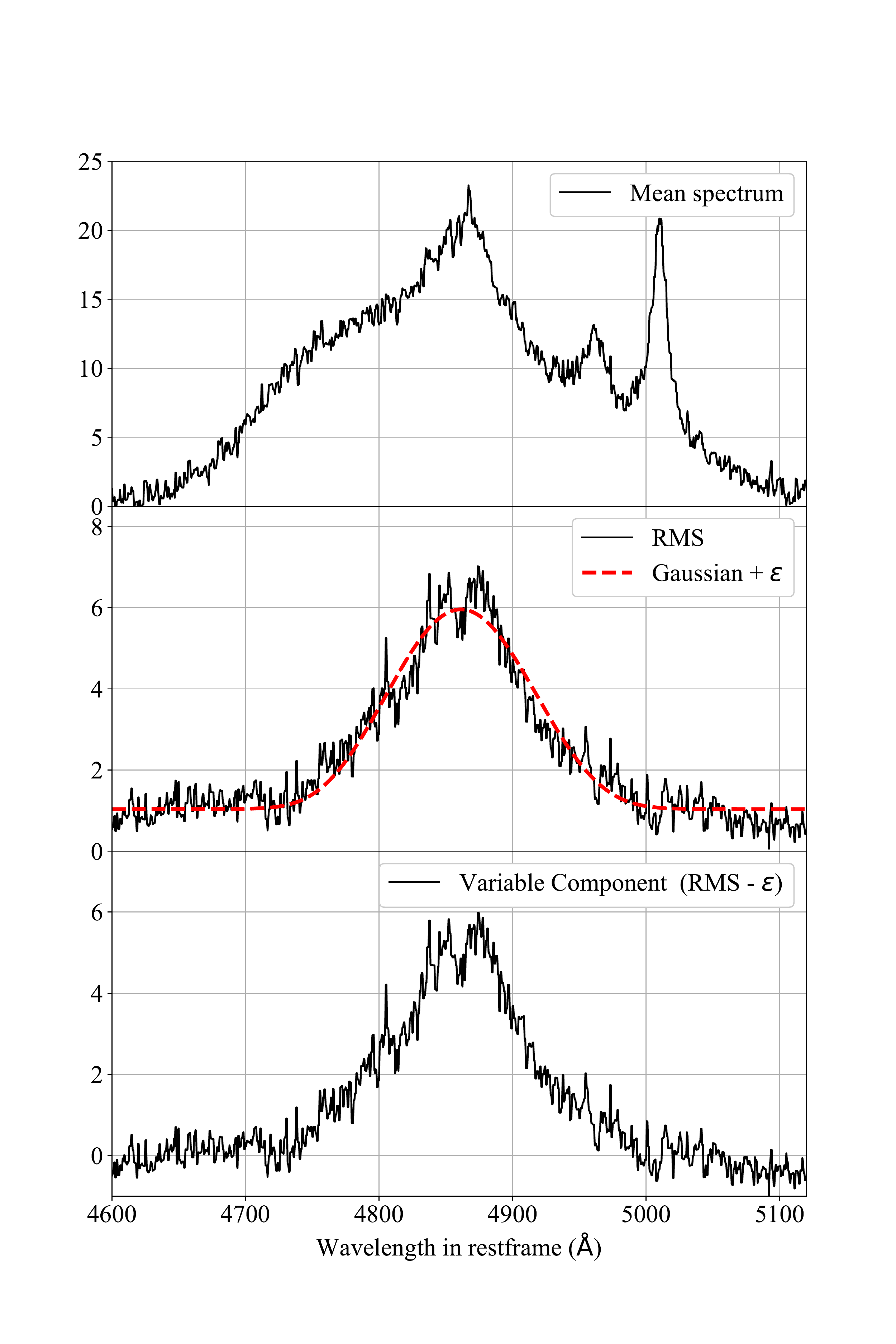}
    \caption{Mean (top), RMS (middle), and the variable component (bottom; baseline subtracted RMS spectrum) spectra of J1258. The middle panel depicts the fitting result for the estimation of $\epsilon$ with Equation~\ref{eq:epsilon_estimation}. The variable component spectrum shows clearly the broad and symmetric $\rm{H\beta}$. The narrow [OIII]4959/5007 lines and asymmetric double peaked component of broad $\rm{H}\beta$ appear only in the mean spectrum, except for weak residuals.}
    \label{fig:mean_RMS}
\end{figure}

Here, it is important to note that the RMS spectrum is affected by the observational errors present in each individual spectrum, in addition to the actual fluctuating component of the BLR. Because the measurement error is difficult to estimate accurately in practice, we assumed in the simplest case that the error is constant within the wavelength range of the BLR. 
In other words, the flux of the RMS ($\equiv f_{\rm RMS}(\lambda)$) was assumed to be expressed in the following equation using the actual variable component ($\equiv f_{\rm variable}(\lambda)$) and observational error ($\equiv \epsilon$).
\begin{equation}
f_{\rm RMS}(\lambda) = (f_{\rm variable}(\lambda)^{2} + \epsilon^{2})^{\frac{1}{2}}
\label{eq:epsilon_estimation}
\end{equation}
To estimate the constant $\epsilon$, we fitted the RMS spectrum with a Gaussian and $\epsilon$ with Equation~\ref{eq:epsilon_estimation} (see the middle panel of Figure~\ref{fig:mean_RMS}). This procedure of removing the measurement error from the RMS spectrum has a significant impact on the calculation of the standard deviation ($\sigma$) of the emission line.

These line spectra (mean, RMS, and variable) are all derived from line spectra with the local continuum already subtracted. The FWHM and $\sigma$ values have been calculated within the wavelength range of 4600 \textrm{\AA} to 5120 \textrm{\AA}. For the mean spectrum, this calculation was performed after subtracting the narrow line components.
The results of measuring the $\sigma$ and FWHM of $\rm{H\beta}$ for the mean spectrum, the RMS spectrum, and the variable component are summarized in Table~\ref{tab:mass_params}.

\begin{table}
	\centering
	\caption{Line width of the mean, RMS, and base reduced RMS spectra. The value of the mean spectrum is measured with narrow component subtracted spectrum}
	\label{tab:mass_params}
	\begin{tabular}{lccc} 
		\hline
		   & mean - narrow & RMS & variable \\
		\hline
            FWHM [km/s]   & $10268 \pm 408$ & $6462 \pm 656$ & $5242 \pm 556$ \\
            $\sigma$ [km/s] & $4727 \pm 13$ & $4119 \pm 79$ & $3429 \pm 157$ \\

		\hline
	\end{tabular}
\end{table}

The mean, RMS, and the variable component spectra for J1258 are shown in Figure~\ref{fig:mean_RMS}. The mean spectrum represents the typical shape of the $\rm{H\beta}$ and $\rm{[OIII]4959/5007}$. One of the characteristics of the mean spectrum is asymmetry (the bluer side is stronger), but the variable component spectrum looks symmetric. The RMS spectrum shows that [OIII]4959/5007 profiles are little variable, which indicates that the flux normalization using [OIII]5007 is working well.

\subsection{Time-lag measurement} \label{sec:time-lag}
To estimate the radius of the BLR and the dust torus, we calculate the time-lag from the continuum light curve for $\rm{H\beta}$, WISE $W1$, and WISE $W2$, respectively. We used the results using two different software algorithms to obtain time-lags: interpolated cross-correlation function \citep[ICCF; ][]{2018ascl.soft05032S, 1998AdSpR..21...57P}, and JAVELIN \citep{2010ascl.soft10007Z, 2011ApJ...735...80Z, 2013MNRAS.431.3319Z}.

ICCF calculates the cross-correlation function with two linearly interpolated light curves. The time-lag and its error are determined by the bootstrapping method; it calculates the cross-correlation function 10,000 times with light curves of different values within error bars, resulting in the distribution of the correlation peak \citep[i.e., the FR/RSS; see ][]{1998PASP..110..660P}. 

On the other hand, JAVELIN assumes a primary light curve to follow the dumped random walk (DRW) process \citep[e.g.,][]{2009ApJ...698..895K} and convolves it with a top-hat transfer function (TF) to make a responding light curve, searching with the Markov chain Monte Carlo method for the most likely values for the following five parameters: the amplitude and time-scale of the DRW process, height and width of the top-hat TF, and response time delay. 

\begin{figure}
	\includegraphics[width=\columnwidth]{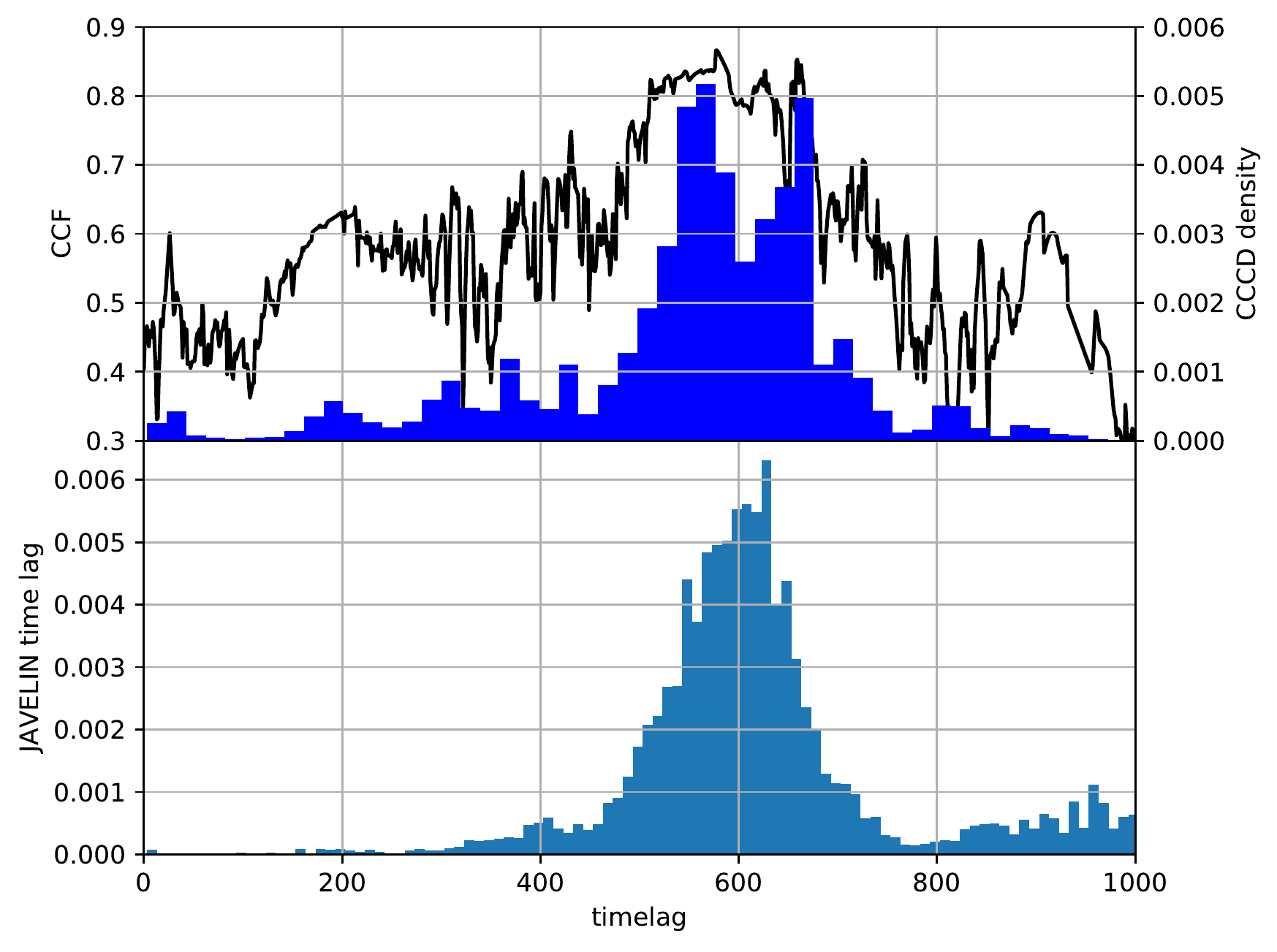}
	\includegraphics[width=\columnwidth]{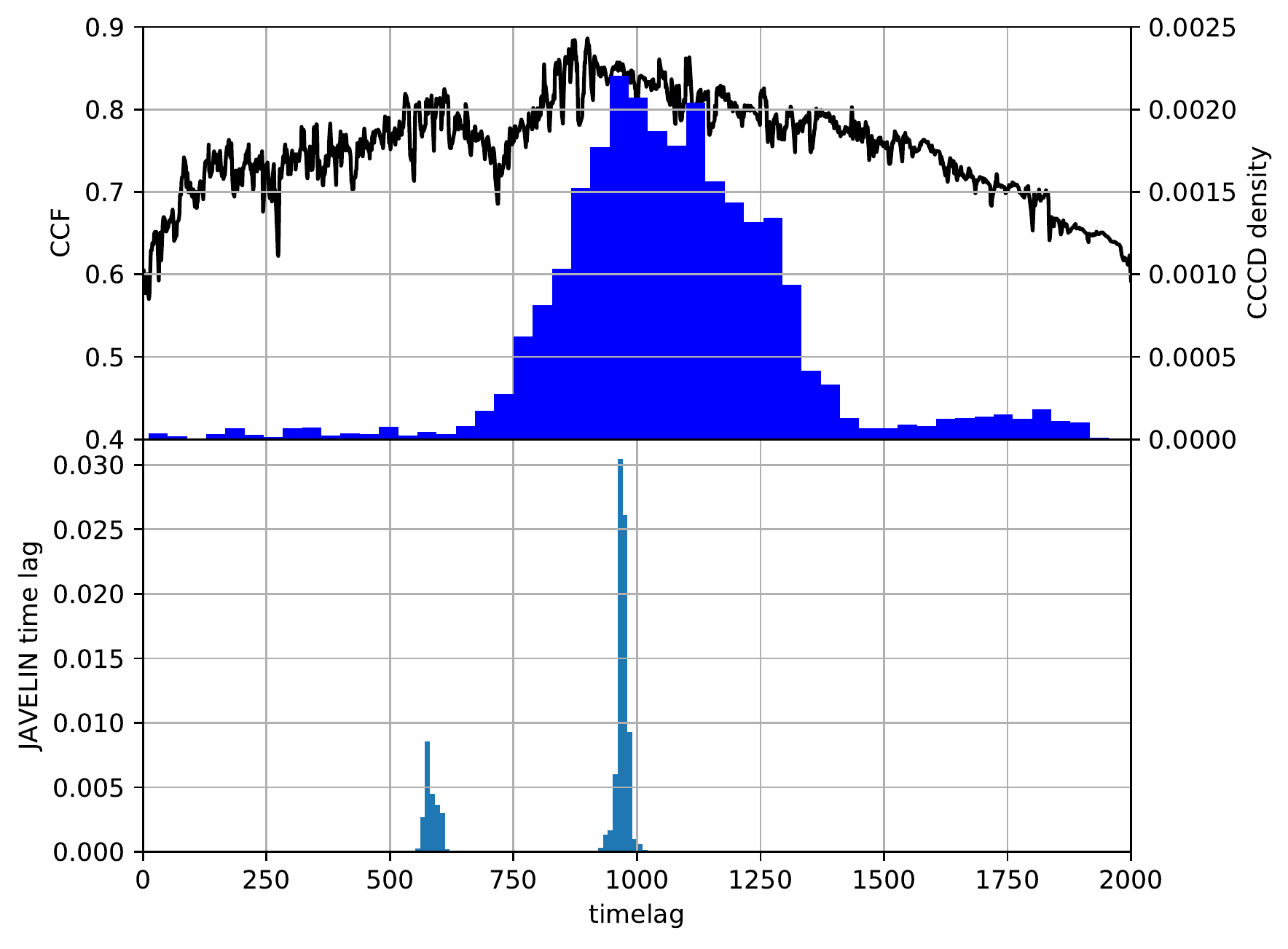}
	\includegraphics[width=\columnwidth]{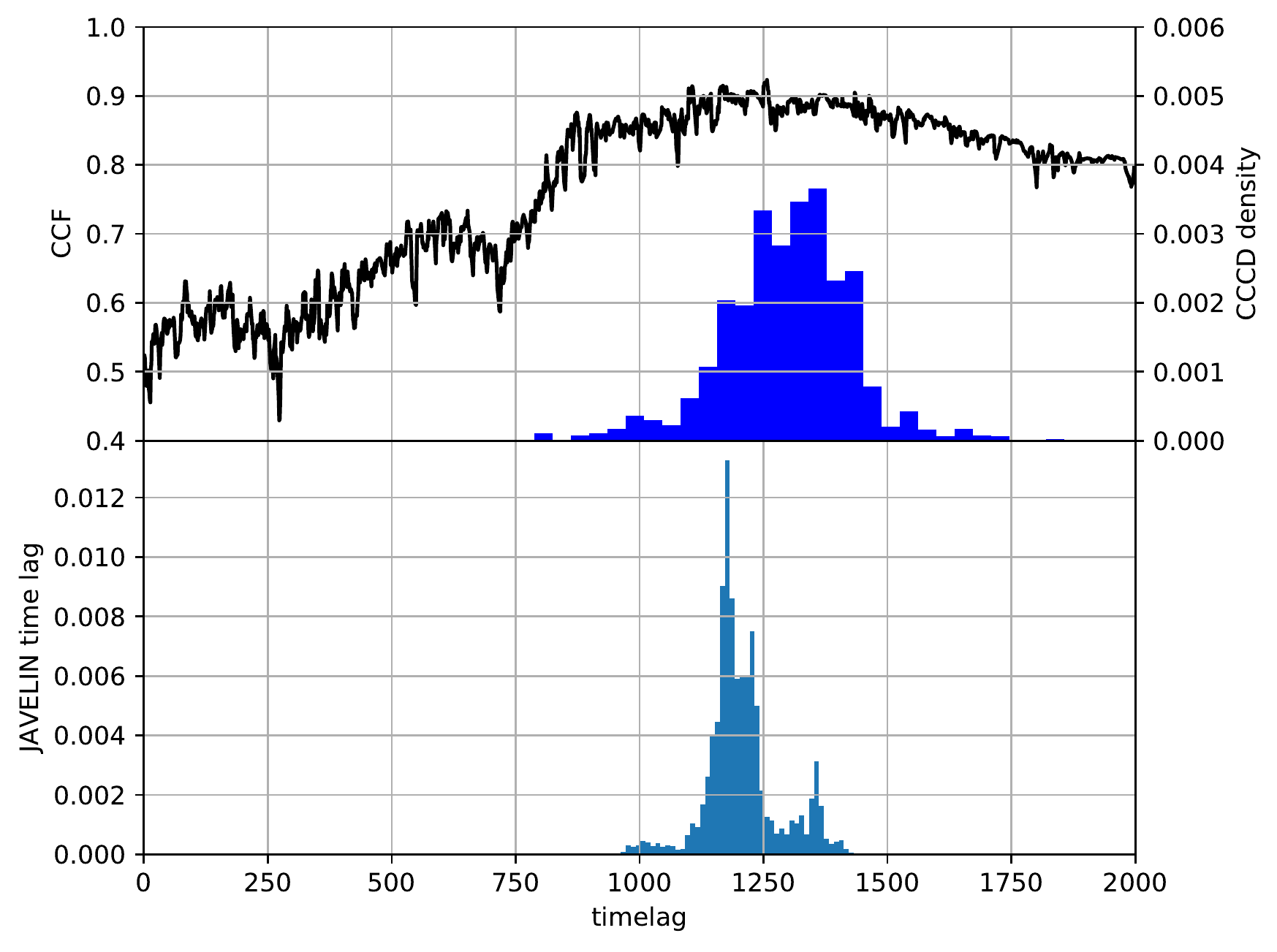}
    \caption{The Cross-Correlation Function and the centroid distribution to the continuum with $\rm{H\beta}$ (top), WISE $W1$ (middle), and WISE $W2$ (bottom) of J1258.}
    \label{fig:hist_ICCCF_javelin}
\end{figure}

The results of time-lag measurements are summarized in Table~\ref{tab:time-lag} and Figure~\ref{fig:hist_ICCCF_javelin}.

\begin{figure}
	\includegraphics[width=\columnwidth]{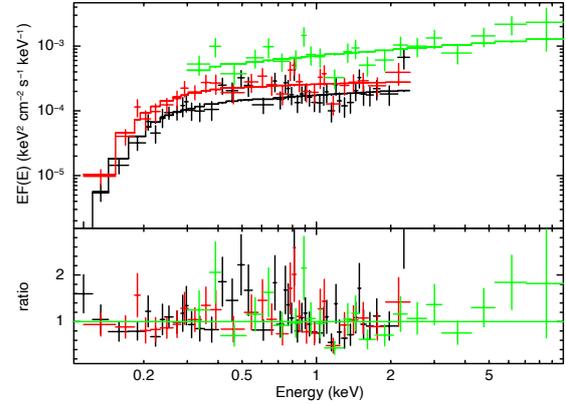}
    \caption{The unfolded X-ray spectra and best-fit models of the observations with ROSAT in 1991 (black), 1992 (red), and Swift/XRT in 2021 (green). The bottom panel shows the ratios between the data and models.}
    \label{fig:x_spectra}
\end{figure}

\begin{table}
	\centering
	\caption{Rest-frame reverberation lags based on ICCF and JAVELIN analysis}
	\label{tab:time-lag}
	\begin{tabular}{lccr} 
		\hline
		Time-lag target & $\tau_{\rm{rest}}$(ICCF) & $\tau_{\rm{rest}}$(JAVELIN) \\
		\hline
		$\rm{H\beta}$ &  $439^{+68}_{-119}$  &  $461^{+60}_{-61}$  \\
            $W1$    &   $805^{+170}_{-137}$     &  $739^{10}_{-291}$   \\
            $W2$    &	$998^{+80}_{-98}$      &  $904^{41}_{-29}$   \\
		\hline
	\end{tabular}
\end{table}

\section{Discussion} \label{sec:Discussion}
In this section, we first estimate the black hole mass. Secondly, we discuss the mechanism of the extreme brightening event of J1258. Finally, based on a review of the information obtained in this study, we discuss the structure of the BLR, and the origin of the BLR.

\subsection{Black Hole Mass Estimation} \label{sec:bh_mass}
In order to discuss the structure of the BLR and the dust torus, we first estimate the mass of the central black hole based on the time-lag information of the BLR. The black hole mass is calculated using Equation~\ref{equation:BHmass}, including the object-specific parameters $f$, $\tau$, and $\Delta V$, which require certain assumptions for each. In this section, we discuss the values of these parameters.

Although the factor $f$ is inherently different for each object because of the influence of BLR structure and viewing angle, previous studies have often applied the same value to different targets as a typical value for many objects. To determine the value of $f$ specifically for each object, a  flexible model fitting with respect to the structure of the BLR would be required. However, in this paper, we consider this to be a subject for future work, and for the time being, adopt the value of $f=4.31\pm1.05$ \citep{2013ApJ...773...90G} from previous research.

The parameter $\tau$ represents the typical radius of the BLR, which is estimated in this paper using two different algorithms, ICCF and JAVELIN. As shown in Table~\ref{tab:time-lag}, there is no significant discrepancy between the results obtained by the two algorithms. Consequently, accepting the results of either algorithm will not significantly affect the following discussion. However, given reports that JAVELIN tends to estimate time lags in a way that fits into observational gaps \citep{2023arXiv230501014S}, here we adopt the results from the ICCF, which involves fewer assumptions.

In this study, we propose a procedure for estimating the $\epsilon$ parameter, which contributes to determining a physically plausible velocity width. 
In Equation~\ref{equation:BHmass}, since $\tau$ represents the response time of the fluctuating component of the emission line, it is considered appropriate for the velocity $\Delta V$ of the emission line to be measured in the variable component of the spectrum. As the RMS spectrum is affected by substantial observational errors, we contend that it is physically more appropriate to estimate the velocity using the variable component, the error-subtracted RMS spectrum, for the calculation of the black hole mass.

In summary, this paper refers the $f$ value from \cite{2013ApJ...773...90G}, and adopts the ICCF estimate for $\tau(439^{+68}_{-119} \ {\rm days})$, and $\sigma (3429 \pm 157 \ {\rm km/s})$ for the variable component of $\Delta V$. It is important to note that regardless of the method employed, there is no difference in the order of magnitude, and no significant impact on the discussion of BLR and dust torus structure in the subsequent sections. In the following analysis, we use $10^{9.64^{+0.11}_{-0.20}}\rm{M_\odot}$ for the black hole mass calculated using Equation~\ref{equation:BHmass}.

\subsection{Mechanism of the extreme variability} \label{sec:Mechanism_variation}
In general, the leading cause of the extreme variation accompanied by CSQs is considered to be intrinsic variability in the mass accretion rate of the disk for the following reasons. First, the variability timescale of BELs and continuum is shorter than the orbital timescale of dust torus ($>\sim 100 \ {\rm yr}$). Second, the reddening in optical \citep[e.g.,][]{2016MNRAS.457..389M} and the X-ray hydrogen column density \citep{2016A&A...593L...9H} did not change before and after the state transition. Third, the brightness in the mid-IR changed in response to continuum emission \citep{2018ApJ...864...27S, 2021PASJ...73..122N}. Finally, polarization was not detected in spectropolarimetry \citep{2017A&A...604L...3H, 2019A&A...625A..54H}. The above results support that CSQs are not explained by any shielding theories, but are consistent with the intrinsic variation of mass accretion rate.

In \cite{2021PASJ...73..122N}, we proposed a hypothesis that the J1258's brightening event was caused by a state transition in the accretion disk, based on its timescale. To reconsider the hypothesis, we performed spectroscopic monitoring and X-ray observations to estimate the optical-to-X-ray spectral indices ($\alpha_{\rm ox}$) in the On/Off state each.

\subsubsection{$\alpha_{\rm{ox}}$ and Eddington ratio}
We analyzed the X-ray spectra of the ROSAT and Swift/XRT data with XSPEC v12.10.1 \citep{1996ASPC..101...17A}.
We consider Galactic absorption by the \textsf{phabs} model and refer the hydrogen column density fixed at the value of \citet{2013MNRAS.431..394W}.
Abundances were assumed at solar values.
Cash statistics \citep{1979ApJ...228..939C} were utilized to fit ROSAT (0.1--2.4~keV) and Swift (0.3--10~keV) spectra, and the spectra were binned to have one count per bin.
The spectra were well reproduced by a single power law model (\textsf{zpowerlw}) with Galactic absorption, as shown in Figure~\ref{fig:x_spectra}.
The best-fitting parameters of observed-frame 2 keV ($\nu F_{\rm 2\,keV}$) and 2--10~keV ($F_{\rm 2-10\,keV}$) fluxes, and rest-frame 2 keV ($\nu L_{\rm 2\,keV}$) and 2--10~keV ($L_{\rm 2-10\,keV}$) luminosities\footnote{The uncertainties of $\nu F_{\rm 2\,keV}$ and $\nu L_{\rm 2\,keV}$ are obtained by the \textsf{cflux} and \textsf{clumin} models in XSPEC. The uncertainties of the 2--10 keV fluxes and luminosities are assumed to the log-scale uncertainties at the 2 keV band.}, all of which are corrected for Galactic absorption, are presented in Table~\ref{tab:x_results}.
The photon indices are $\sim$1.7--1.9, consistent with a typical value in AGNs \citep[e.g.,][]{2014ApJ...786..104U,2017ApJS..233...17R}.
We find that the rest-frame 2~keV luminosities increase from $\sim 1 \times 10^{44}$~erg/s in 1991 and 1992 to $\sim 4 \times 10^{44}$~erg/s in 2021.

\begin{table}
	\centering
	\caption{Best-fit X-ray spectral parameters, observed-frame absorption-corrected fluxes, rest-frame luminosities, and reduced C statistics}
	\label{tab:x_results}
	\begin{tabular}{lccc} 
		\hline
		Parameters & ROSAT & ROSAT & Swift \\
		& (1991) & (1992) & (2021) \\
		\hline

        Photon Index & $1.82^{+0.06}_{-0.05}$ & $1.94^{+0.07}_{-0.08}$ & $1.73^{+0.10}_{-0.10}$ \\
        $\nu F_{\rm 2\,keV}$ [10$^{-13}$ erg/s/cm$^2$] & $3.25^{+0.29}_{-0.28}$ & $4.44^{+0.54}_{-0.49}$ & $14.07^{+1.18}_{-1.13}$ \\
        $F_{\rm 2-10\,keV}$ [10$^{-13}$ erg/s/cm$^2$] & $6.03^{+0.54}_{-0.51}$ & $7.52^{+0.91}_{-0.83}$ & $28.27^{+2.37}_{-2.27}$ \\
        $\nu L_{\rm 2\,keV}$ [10$^{44}$ erg/s] & $0.96^{+0.07}_{-0.07}$ & $1.36^{+0.13}_{-0.13}$ & $4.07^{+0.29}_{-0.29}$ \\
        $L_{\rm 2-10\,keV}$ [10$^{44}$ erg/s] & $1.78^{+0.14}_{-0.13}$ & $2.29^{+0.24}_{-0.22}$ & $8.16^{+0.59}_{-0.57}$ \\
        C/dof & 166.8/174 & 147.2/140 & 147.8/192 \\

  \hline
	\end{tabular}
\end{table}

We compare the Eddington ratio and $\alpha_{\rm{ox}}$ of before and after the state transition with the similar method of \cite{2021ApJ...912...20J}. $\alpha_{\rm{ox}}$ is a spectral index between optical and X-ray characterizing the state of the accretion disk. We use the following definition to calculate $\alpha_{\rm{ox}}$ \citep{1979ApJ...234L...9T},
\begin{equation}
    \alpha_{\rm{ox}} = - \frac{{\rm log}(\lambda L_{2500\,\text{\AA}}) - {\rm log}(\nu L_{\rm{2\,keV}})}{{\rm log}\nu_{2500\,\text{\AA}} - {\rm log}\nu_{\rm{2\,keV}}} + 1
	\label{eq:alpha_ox}
\end{equation}
in which the continuum luminosity at 2500 $\text{\AA}$ is estimated by extrapolation of the fitted power-law component. For the spectral fitting, the power-law component, FeII template \citep{1992ApJS...80..109B}, and the Balmer continuum component were applied to 4450--4500 and 5150--5250 $\text{\AA}$ avoiding narrow emission lines and BELs. The errors were estimated by Monte Carlo simulation. The bolometric luminosity is estimated using the correction formula in \cite{2010A&A...512A..34L}. 
\begin{equation}
    {\rm log}(L_{\rm{bol}}) = {\rm log}(L_{2-10\,\rm{keV}}) + 15.61 - 1.853\alpha_{\rm{ox}} + 1.226\alpha_{\rm{ox}}^2 
	\label{eq:Lbol}
\end{equation}
The obtained $\alpha_{\rm{ox}}$, bolometric luminosity, and Eddington's ratio before and after the state transitions are summarized in Table~\ref{tab:Comparison}.

\begin{table}
	\centering
	\caption{
 Table summarizing the parameters representing the accretion disk state before and after state transitions. Each parameter is defined as follows: ${\rm log}(\lambda L_{2500\,\text{\AA}})$ represents the $\lambda L$ at $2500~\text{\AA}$, $L_{\rm bol}$ is the estimated bolometric luminosity using Equation~\ref{eq:Lbol}, the Eddington ratio is calculated from $L_{\rm{bol}}$ and the black hole mass, and $\alpha_{\rm{ox}}$ is calculated using Equation~\ref{eq:alpha_ox}.
    }
	\label{tab:Comparison}
	\begin{tabular}{lccr} 
		\hline
		Parameters & Off State & On State \\
		\hline
		${\rm log}(\lambda L_{2500 \text{\AA}})$ & $44.289\pm0.010$ & $45.337\pm0.001$ \\
		$\alpha_{\rm{ox}}$ & $1.02\pm0.07$ & $1.29\pm0.05$\\
		${\rm log}L_{\rm{bol}}$ & $45.3\pm0.22$ &  $46.1\pm0.11$\\
		Eddington Ratio & $0.004\pm0.009$ & $0.024\pm0.051$\\
	
		\hline
	\end{tabular}
\end{table}

\begin{figure}
    \includegraphics[width=\columnwidth]{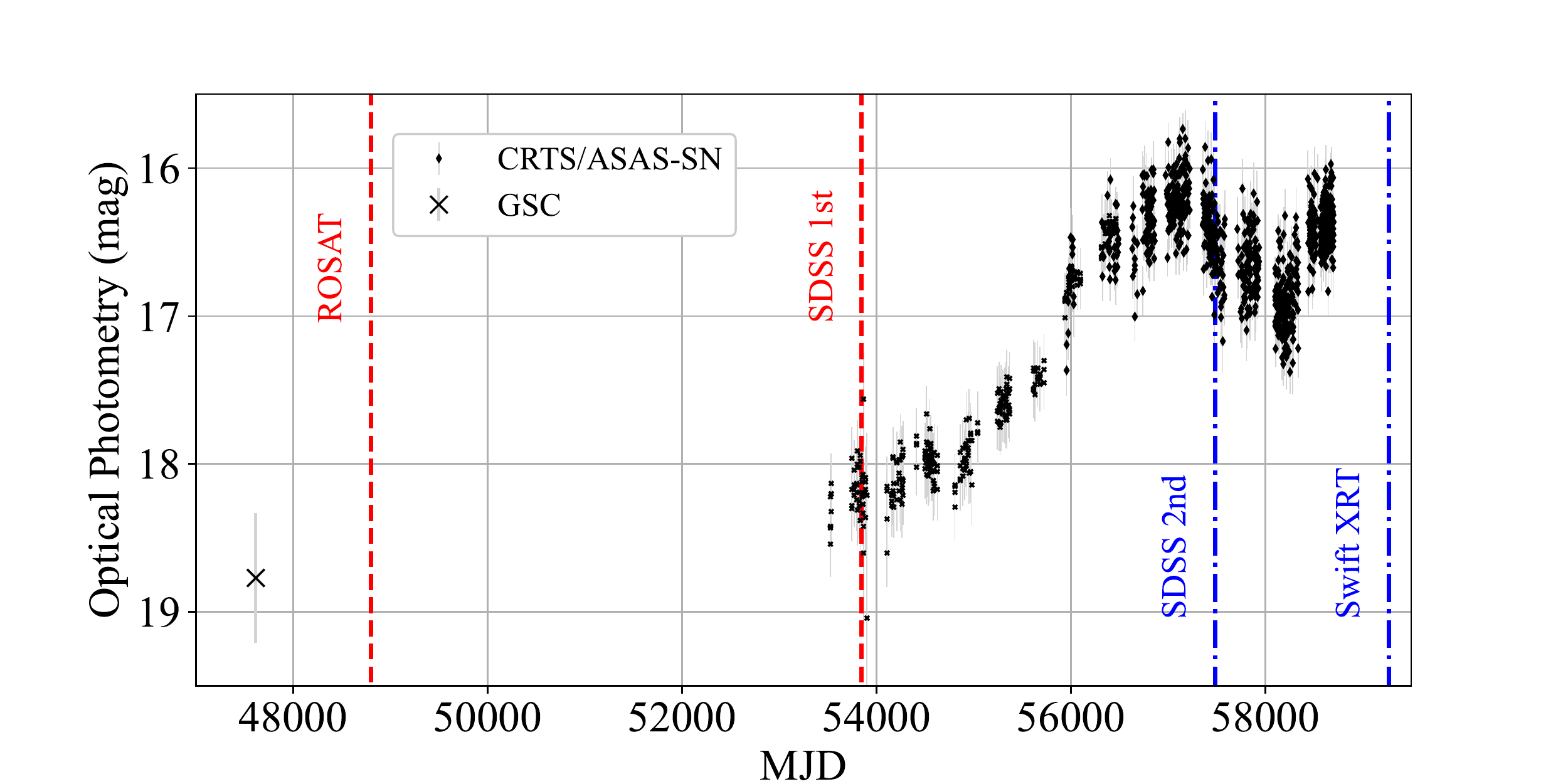}
    \caption{The optical light curve with periods of optical spectra and X-ray observation. The black dots represent the optical light curve, and the vertical lines drawn with dotted lines represent the time of spectral or X-ray observations. In addition to the CRTS and ASAS-SN light curves, data from The Guide Star Catalogue \citep{2008AJ....136..735L} are plotted. We interpret the data before MJD = 54000 are considered as Off state, and the data after MJD = 57000 as On state.}
    \label{fig:on_off_lc}
\end{figure}

We used optical spectra and X-ray data taken during Off and On states each. Since the optical light curve in Figure~\ref{fig:on_off_lc} shows that the state transition happened in 2007 (MJD around 54000) to 2015 (MJD around 57000), we consider the status before 2007 as Off state, while the status after 2015 as On state.

\cite{2021ApJ...912...20J} compared the $\alpha_{\rm{ox}}$ of 10 previously discovered CSQs with the simulated results from \citet{2011MNRAS.413.2259S} based on X-ray observations of the X-ray binary GRO J1655-40 extended to AGNs. They confirmed that it is consistent with the prediction of the state transition model.
\cite{2021ApJ...912...20J} found the negative correlation between Eddington ratio and $\alpha_{\rm{ox}}$ for the objects with low (< 0.01) Eddington ratio, while \cite{2010A&A...512A..34L} reported the positive correlation for the objects with high (> 0.01) Eddington ratio. 

\subsubsection{Accretion rate variation}
\begin{figure}
    \includegraphics[width=\columnwidth]{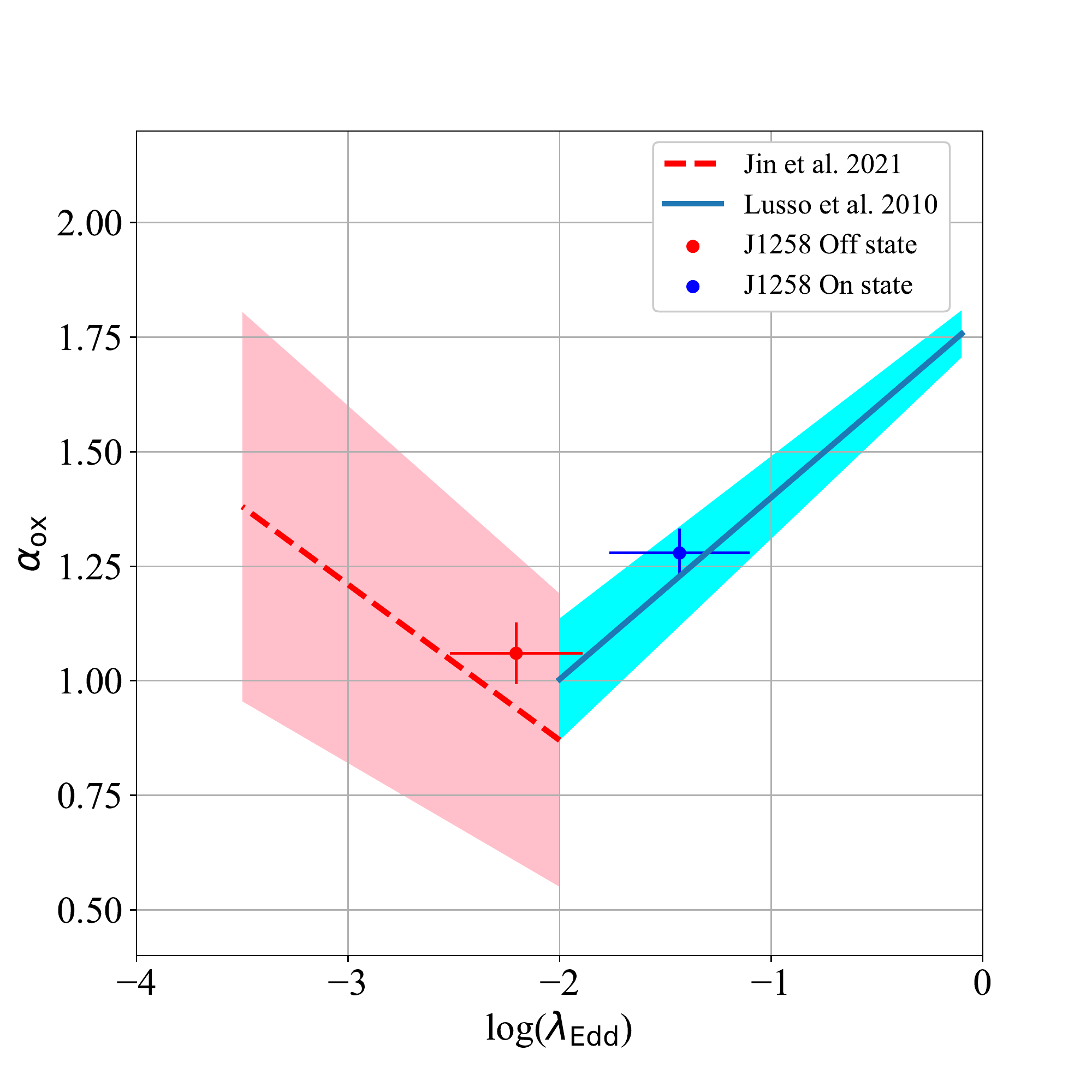}
    \caption{
    Comparison of the measured $\alpha_{\rm{ox}}$ and $\lambda_{\rm{Edd}}$ for J1258 in its On and Off states. The red dashed line and surrounding red region indicate the fitting results and associated errors for the Off state derived from quasars and CSQs, as referenced by \citep{2021ApJ...912...20J}. In contrast, the blue solid line and surrounding the blue region represent the $\alpha_{\rm{ox}}$ - $\lambda_{\rm{Edd}}$ relationship and its associated errors for the On state, based on findings from the XMM-COSMOS survey \citep{2010A&A...512A..34L}.
    }
    \label{fig:edd_alpha}
\end{figure}

The measured Eddington ratio and $\alpha_{\rm{ox}}$ of J1258 in the Off and On state each are consistent with the state transition model suggested in the previous research \citep{2018MNRAS.480.3898N}. We show the comparison of the Eddington ratio and $\alpha_{\rm{ox}}$ on Figure~\ref{fig:edd_alpha} together with the best-fit results of \cite{2021ApJ...912...20J} and \cite{2010A&A...512A..34L}. 

One of the possible causes of the state transition is thermal disk instability \citep{2018MNRAS.480.3898N}. According to the disk instability model, the accretion disk forms a geometrically thin disk at high Eddington ratios \citep{1973A&A....24..337S} characterized by soft excess in X-ray, whose accretion state is called the high/soft state. During the low Eddington ratio, the inner region of the thin disk is evaporating \citep{1997ApJ...489..865E}, resulting in a hot and radiatively inefficient accretion flow \citep{1976ApJ...204..187S}, possibly advection dominated \citep{1994ApJ...428L..13N}. In this state, the X-ray spectrum is dominated by Compton scattering of hard X-rays originating from the accretion flow, called the low/hard state. The spectral index from optical to soft X-ray component, $\alpha_{\rm{ox}}$, is used to characterize these spectral states. The correlation between $\alpha_{\rm{ox}}$ and the Eddington ratio is predicted theoretically to change according to certain rules (negatively correlated to the Eddington ratio during low/hard state, while positively correlated during high/soft state) during the state transition, and has been confirmed observationally in X-ray binaries \citep[e.g.,][]{2003MNRAS.338..189M, 2010A&A...520A..98D}.

Another possibility is that the state transition was triggered by external factors such as the tidal disruption of a star \citep{2015MNRAS.452...69M, 2020ApJ...898L...1R}. The outburst caused by tidal disruption events (TDEs) is expected to show a decline in flux with $t^{-5/3}$ \citep{2020SSRv..216..124V}. In the case of J1258, however, it shows brightening again from around MJD = 58000 (Figure~\ref{fig:on_off_lc}), which is inconsistent with the expected patterns of the light curve of TDEs.

One other possibility is that J1258 is a pair of SMBHs in a binary system. Asymmetric BELs are often referred to as one of the characteristics of binary SMBHs \citep[][]{2012ApJS..201...23E, 2015ApJS..221....7R}. That is a situation where two SMBHs each form an AGN with their own BLRs. Given this situation, if a state transition occurs in one of the AGNs there is no need for occurring the state transition in the other AGN. Furthermore, if the interaction of the two SMBHs efficiently extracted angular momentum, it could be considered a trigger for the large-scale brightening event \citep{2020A&A...643L...9W}. To confirm this hypothesis, it would be helpful to test whether the velocity shift in the asymmetric component of the BELs changes according to the orbital motion. However, it is not easy to calculate the velocity shift of BELs independently of varying intensities because the quality of the obtained spectra in this study is not high enough. If a velocity shift were to present, it would be valid to compare the spectra again after sufficient time has passed, which could be verified in future studies.

We conclude here the main cause of the J1258 brightening event is a variation of the intrinsic accretion power caused by the state transition of the accretion disk, as we mentioned in \cite{2021PASJ...73..122N}.

\subsection{Structure of the BLR} \label{sec:Structure_Origin_BLR}
\begin{figure}
    \includegraphics[width=\columnwidth]{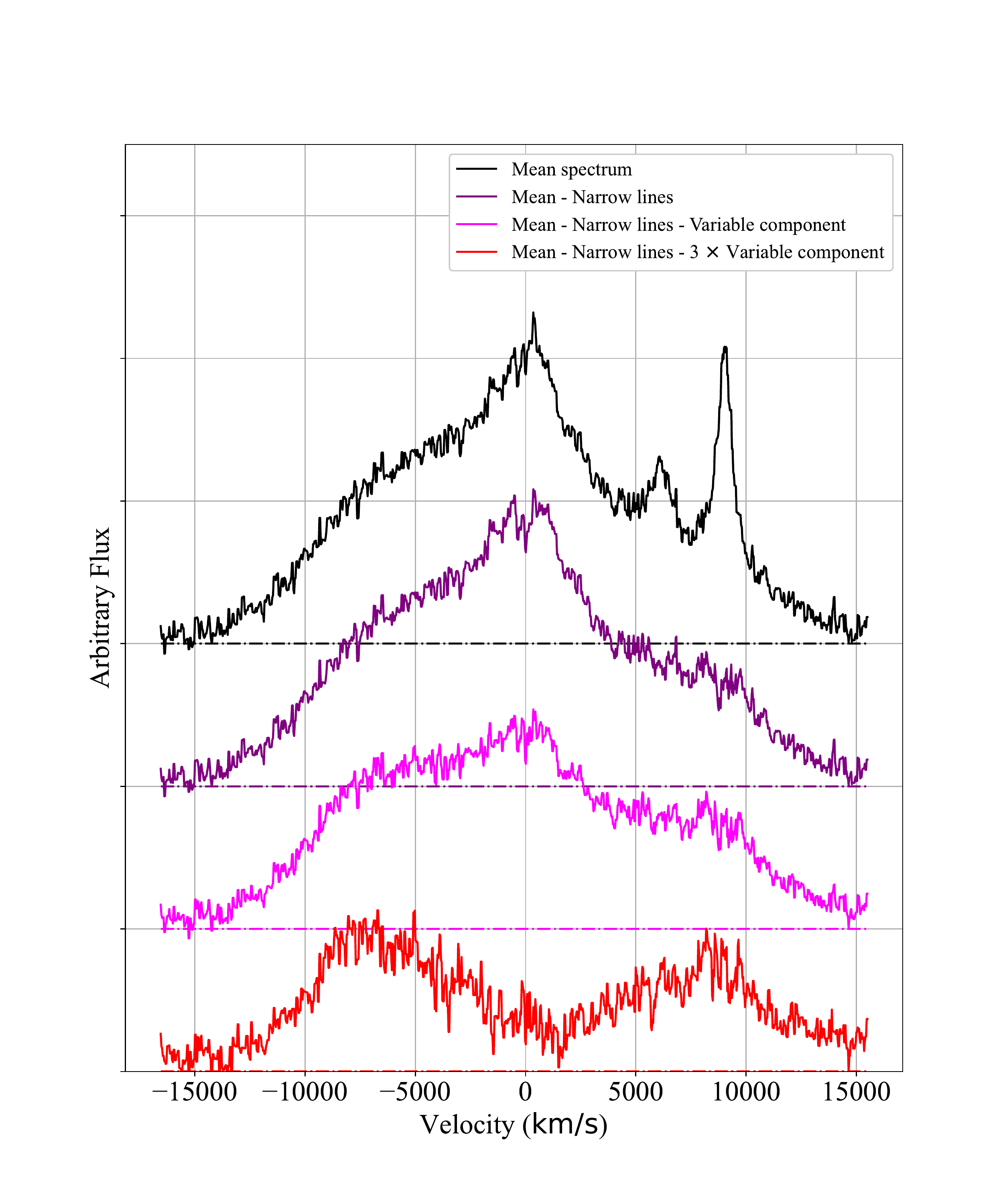}
    \caption{Diagram visualizing the double-peaked component on the BEL. The top spectrum is the mean spectrum, the second from the top is the mean spectrum minus the narrow emission line, and the third is the second spectrum minus the variable component. The bottom spectrum is the mean spectrum minus three times the variable component. The baseline of each spectrum has been added to a constant for clarity.}
    \label{fig:mean_minus_rms}
\end{figure}

We discuss the structure and the origin of the BLR using the extreme variation of the BEL of J1258. The spectra in Figure~{\ref{fig:mean_RMS}} show the mean, the RMS, and the variable component of $\rm H\beta$, respectively. Examining Figure~{\ref{fig:mean_RMS}}, we can see that the shapes of the mean and variable spectra are very different. To show the differences better, we depict the mean spectrum with subtracted narrow lines in Figure~\ref{fig:mean_minus_rms}. This figure also shows how the BEL shape would look like if there were greater changes in the variable component.
Here, we simply assume that the line shape changes with a constant multiple of the variable component in all velocities. The spectrum on the top of Figure~{\ref{fig:mean_minus_rms}} represents the mean spectrum. The second spectrum from the top is the mean spectrum that the narrow lines (the narrow components in $\rm{H\beta}$ and [OIII]4959/5007) are subtracted. 
The third and fourth spectra from the top are 1 or 3 times of the variable component subtracted from the second spectrum (the narrow lines subtracted mean spectrum).
We can see that the spectral shape at the bottom of Figure~{\ref{fig:mean_minus_rms}} shows a double-peaked profile. On the other hand, the variable component has a relatively symmetrical, single-peaked shape. These represent that the spectra typically have a double-peaked component, but the single-peaked component changes with continuum variation, indicating that there are two distinct regions in the BLR \citep{2020ApJ...905...75H}.

\begin{figure}
    \includegraphics[width=\columnwidth]{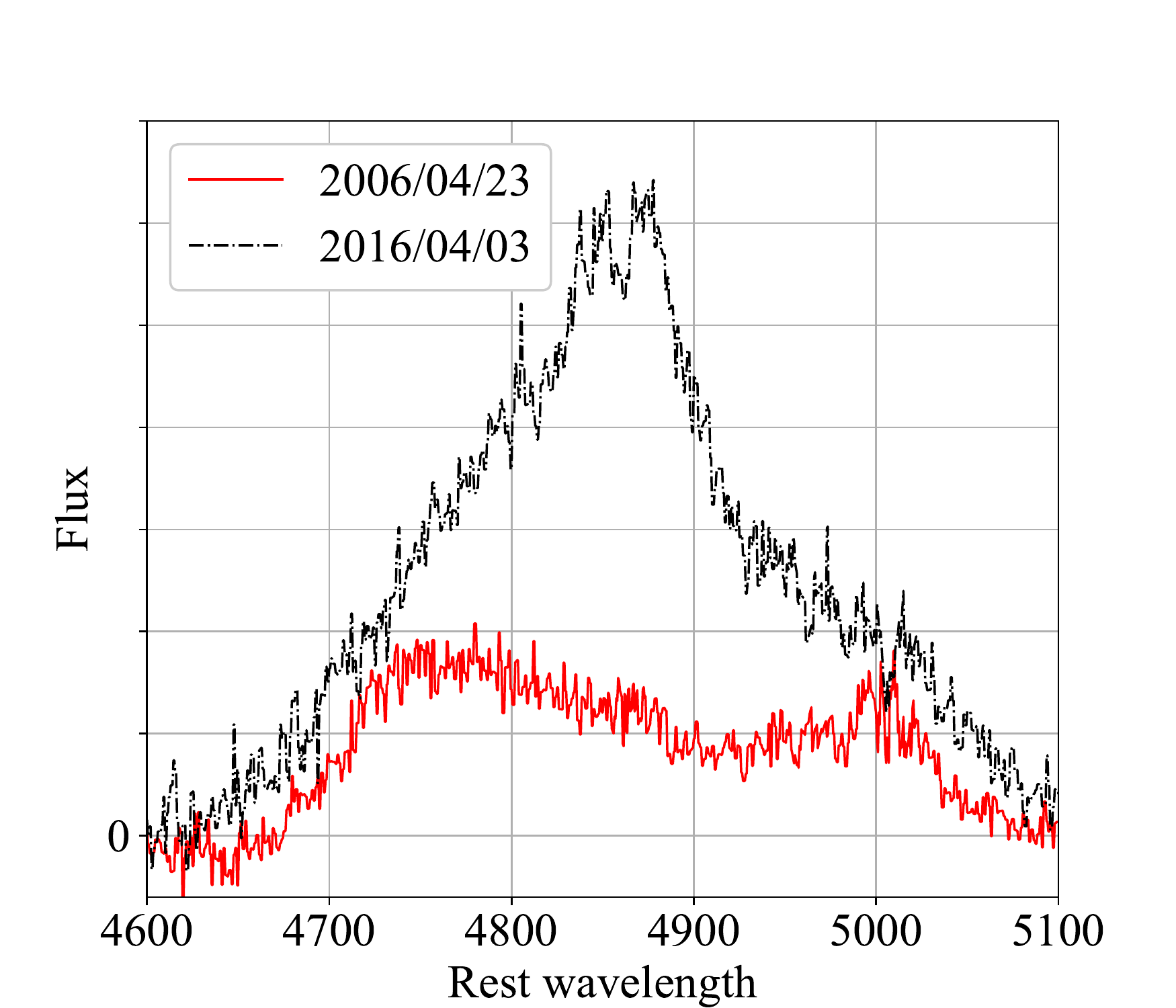}
    \caption{Narrow lines subtracted $\rm H\beta$ from SDSS. The black spectrum was taken on 2006/04/23, and the red spectrum was taken on 2016/04/03.}
    \label{fig:SDSS_Hb}
\end{figure}

Indeed, since this object was observed during a relatively faint period, the double-peaked shape can be seen in real data. Figure~\ref{fig:SDSS_Hb} compares the $\rm H\beta$ shape by subtracting the continuum and narrow emission lines ($\rm H\beta$ and [OIII]4959/5007) based on the spectra from SDSS. The spectrum observed in 2006 has a shape with a small peak at the central wavelength as well as peaks on each side. Considering the central component as variable, we can see a double-peaked component less responsive to changes in the continuum luminosity. Using J1258's extreme luminosity variation, we successfully clarified the existence of components with different characteristics in the BEL.

Based on the above spectral components, the BLR of J1258 is expected to consist of two different regions emitting a single-peaked component (RMS spectrum) and a double-peaked component (the bottom spectrum in Figure~\ref{fig:mean_minus_rms}), respectively. Here we name these components Region~X and Region~Y, respectively:

\begin{itemize}
    \setlength{\leftskip}{1.0cm}
    \setlength{\itemsep}{1mm} 
    \item[\textbf{Region~X:} ] Emitting a relatively slow ($\sim$3,000 km/s), single-peaked line that responds to the state transition of the accretion disk.
    \item[\textbf{Region~Y:} ] Emitting a relatively fast ($\sim$7,000 km/s), double-peaked line, which is not likely affected by the state transition of the accretion disk.
\end{itemize}

Region~X has the same characteristics as the BLR seen in typical quasars, whose size can be estimated from the measured time-lag of the reverberation observation. The BLR size of Region~X was found to be 0.37 pc (439 light days) from the center (Section~\ref{sec:Analysis_Result}).

The kinematics responsible for the double-peaked emission lines observed in Region Y could potentially be attributed to fast outflow, fast inflow, or rotating ring-like structures. While strong outflows and inflows are typically influenced by the mass accretion rate, as suggested in studies such as \citet{2014MNRAS.438.3340E}, our observations indicate that the double-peaked component of Region Y did not exhibit significant variations during state transitions (see Figure~\ref{fig:mean_RMS}). This observation suggests that these emissions might occur independently of the accretion disk's activities, implying a more stable and passive kinematic process. Although the lack of response to continuum variation does not entirely rule out the possibility of outflows or inflows, it encourages consideration of a scenario where these processes are steady and unrelated to the accretion disk's dynamics. While acknowledging that outflow and inflow cannot be completely dismissed, the evidence aligns more consistently with a rotating ring-like structure as the source of the double-peaked broad $\rm H\beta$ emission, as discussed in previous research (e.g., \citet{1994ApJS...90....1E, 2003ApJ...599..886E}). This interpretation is considered more plausible in our context, given the observed stability of the emission lines, though further investigation is warranted to elucidate these kinematics fully.

\begin{figure}
    \includegraphics[width=\columnwidth]{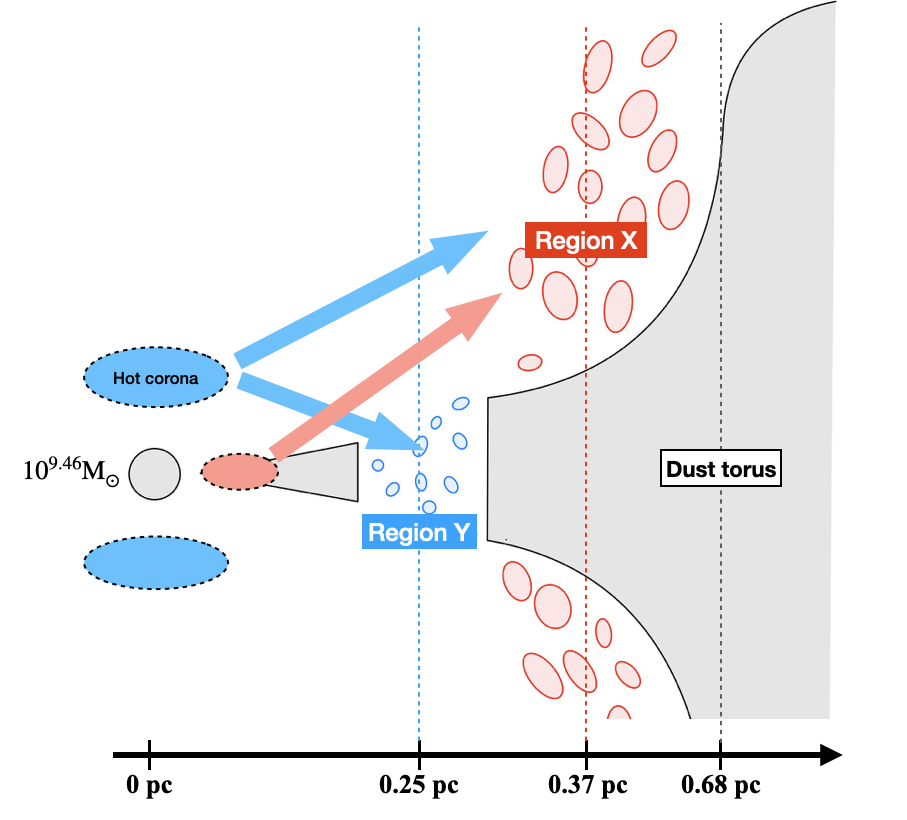}
    \caption{
    Conceptual view of the central core structure of J1258 in On state inferred from the time-lag and wavelength shift. The distance to Region~X and the black hole mass are estimated from the $\rm{H\beta}$ time-lag. The distance to Region~Y is estimated from the determined black hole mass, and the distance to the inner edge of the dust torus is estimated from the WISE $W1$ time-lag.
    }
    \label{fig:BLR_structure}
\end{figure}

With the above points taken into account, it is possible to combine them into a single figure without any geometrical inconsistency. The distance to Region~X can be determined from the time-lag, which is about 0.37 pc. The distance to Region~Y is about $0.29\ \rm{pc} \sim 1400\ \rm{R_g} (\equiv \frac{GM}{c^2})$, which can be explained by the rotation of the disk, given that the black hole mass is $10^{9.64^{+0.11}_{-0.20}}\rm{M_\odot}$ and the typical speed is about 8,000 km/s (the peak velocity of the double-peaked component). If the asymmetric shape is caused by a bias in the distribution of ionized gas in Region~Y, the period over which the gas makes one cycle with Kepler motion is about 225 years. Therefore, detecting a velocity shift over a 20-year observation period would be difficult. The distance to the dust torus is estimated by the response time of $W1$ to be 0.68 pc (805 light days). These locations are shown in Figure~\ref{fig:BLR_structure}.

Assuming the structure of Figure~\ref{fig:BLR_structure}, it is also possible to explain why Region~Y was not affected by the state transition of the accretion disk. The brightening phenomenon due to the state transition from the low/hard state to the high/soft state is caused by the variation of soft X-rays from the inner part of the accretion disk being larger than that of hard X-rays from the hot corona ($T \sim 10^{9} \ \rm{K}$) region \citep{2018MNRAS.480.3898N}. That is, the intensity of the hard X-ray itself does not change much before and after the state transition, which is also confirmed observationally \citep[e.g.,][]{2018ApJ...866..123M}. 
The Compton scattering component from the coronal region ranges from X-ray to the UV which contributes to the ionization of the Balmer lines.
Suppose that Region~Y is located right next to the disk so that only the axially extended hot corona region can supply ionizing photons to Region~Y. In that case, this is a possible explanation for why the intensity of the emission lines emitted from Region~Y is less affected by the state transition.

\begin{figure}
    \includegraphics[width=\columnwidth]{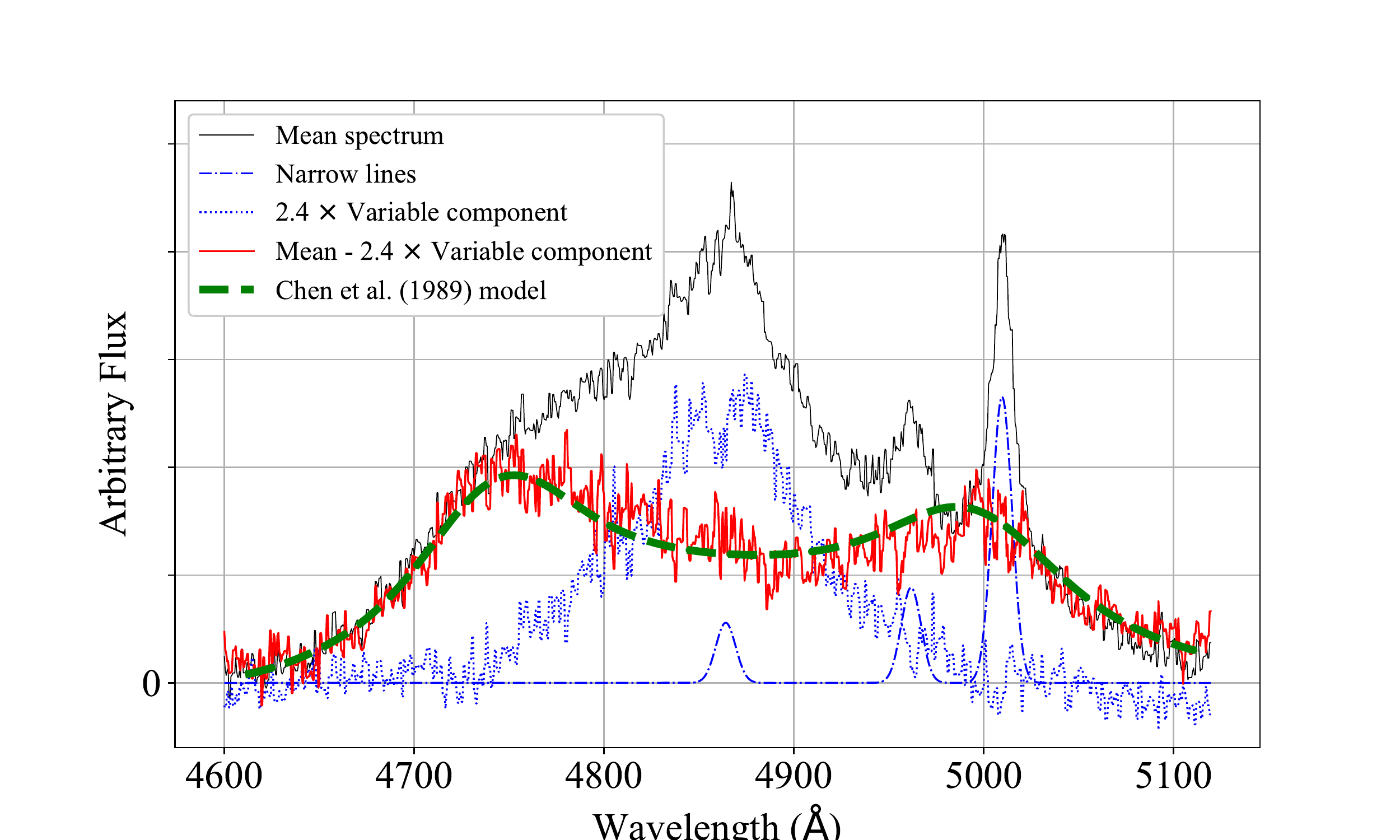}
    \caption{
    Example of reproduction of Region~Y component spectrum by a model of \citet{1989ApJ...339..742C}. The solid black line is the mean spectrum, which is decomposed into a narrow emission line (blue dash-dot line), a constant multiple of the variable component (blue dot line), and Region~Y component (solid red line). The green dotted line shows an example of fitting the Region~Y component with the model of \citet{1989ApJ...339..742C}. }
    \label{fig:Hbeta_modelfitting}
\end{figure}

We confirmed that the double-peaked feature of Region~Y can be reproduced by the structure shown in Figure~\ref{fig:BLR_structure} with model calculation. We adopt a model of \cite{1989ApJ...339..742C} here. This model can calculate the BEL spectrum, including relativistic effects, for a BLR distributed as a thin Keplerian disk. This model is defined by five parameters; the viewing angle $i$, the inner edge of the BLR $\xi_1$, the outer edge of the BLR $\xi_2$, the emissivity index $q$, and the local broadening parameter $\sigma_l$. The model assumes that the surface intensity distribution of the disk is determined by Equation~\ref{eq:emissivity}, which is a function of $\xi_1$ and $\xi_2$ and $q$, and that locally emission lines are emitted with a velocity component of Keplerian rotation and a velocity dispersion of $\sigma_l$. 
The surface intensity $\epsilon(\xi)$ is defined as follows:
\begin{equation}
    \epsilon (\xi) = \frac{\epsilon_{0}}{4\pi}\xi^{-q} {\rm \ [ergs \ cm^{-2} \ s^{-1} \ sr^{-1}]},
    \label{eq:emissivity}
\end{equation}
where, $\epsilon_0$ is a peak value of the emissivity and $\xi$ is radius from the center of the disk.
We show here as an example of parameters that can reproduce the qualitative features of Region~Y, the results calculated with $i=47^{\circ}$, $\xi1=150 R_s$($R_s$ is Schwarzschild radius; $150 R_s = 0.04 \ {\rm pc}$), $\xi_2=900 R_s \ (= 0.25 \ {\rm pc})$, $q=0.9$, and $\sigma_l=1700 \ {\rm km/s}$ are shown in Figure~\ref{fig:Hbeta_modelfitting}, where the spectral shape obtained by subtracting 2.4 times the variable component from the mean spectrum is well reproduced. This result shows that the observed spectral shape of the BEL can be qualitatively reproduced by considering a disk-shaped BLR (Region~Y) with a thin hole inside the BLR radius (at Region~X) estimated by the time delay, as shown in Figure~\ref{fig:BLR_structure}.

\begin{figure}
    \includegraphics[width=0.9\columnwidth]{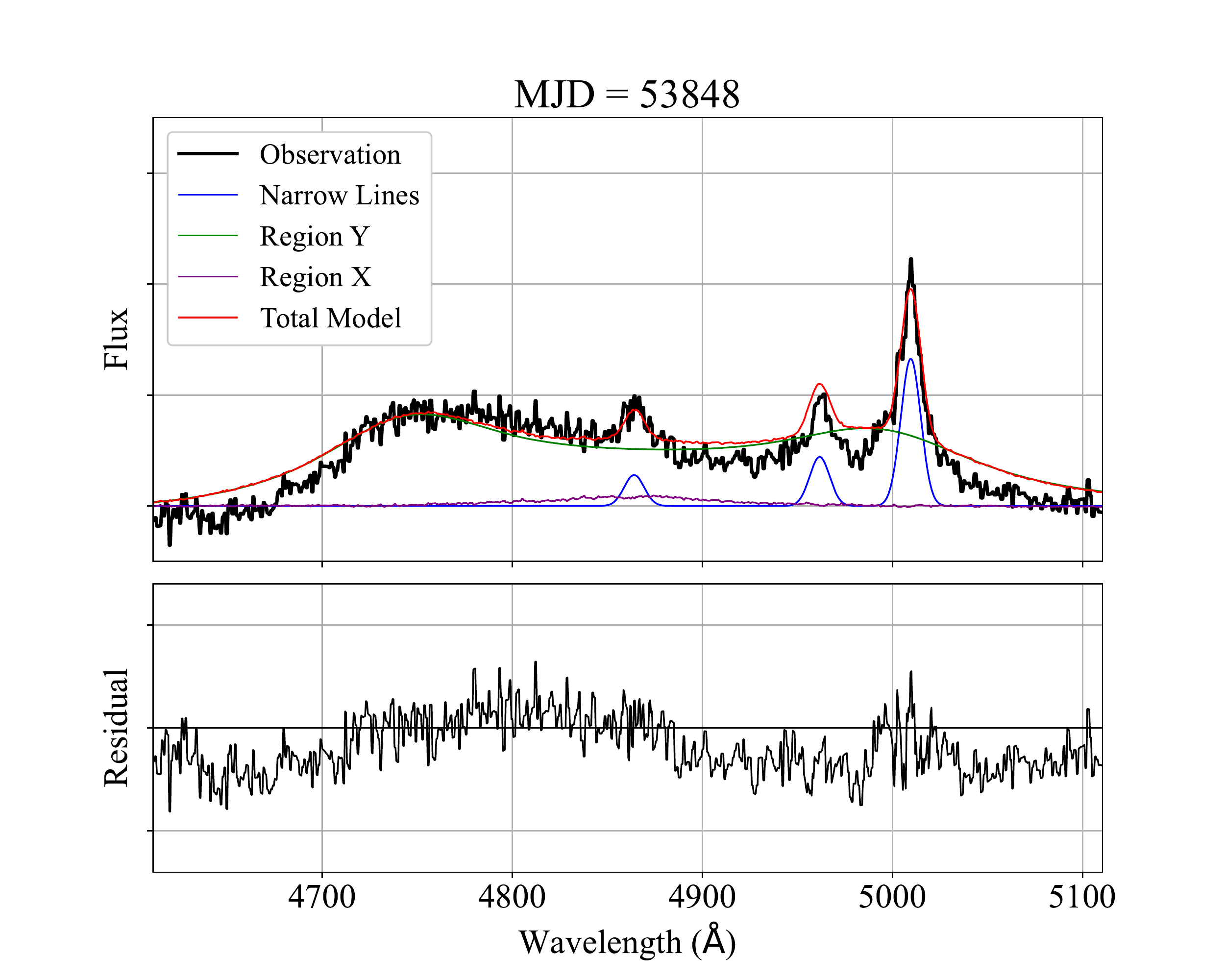}
    \includegraphics[width=0.9\columnwidth]{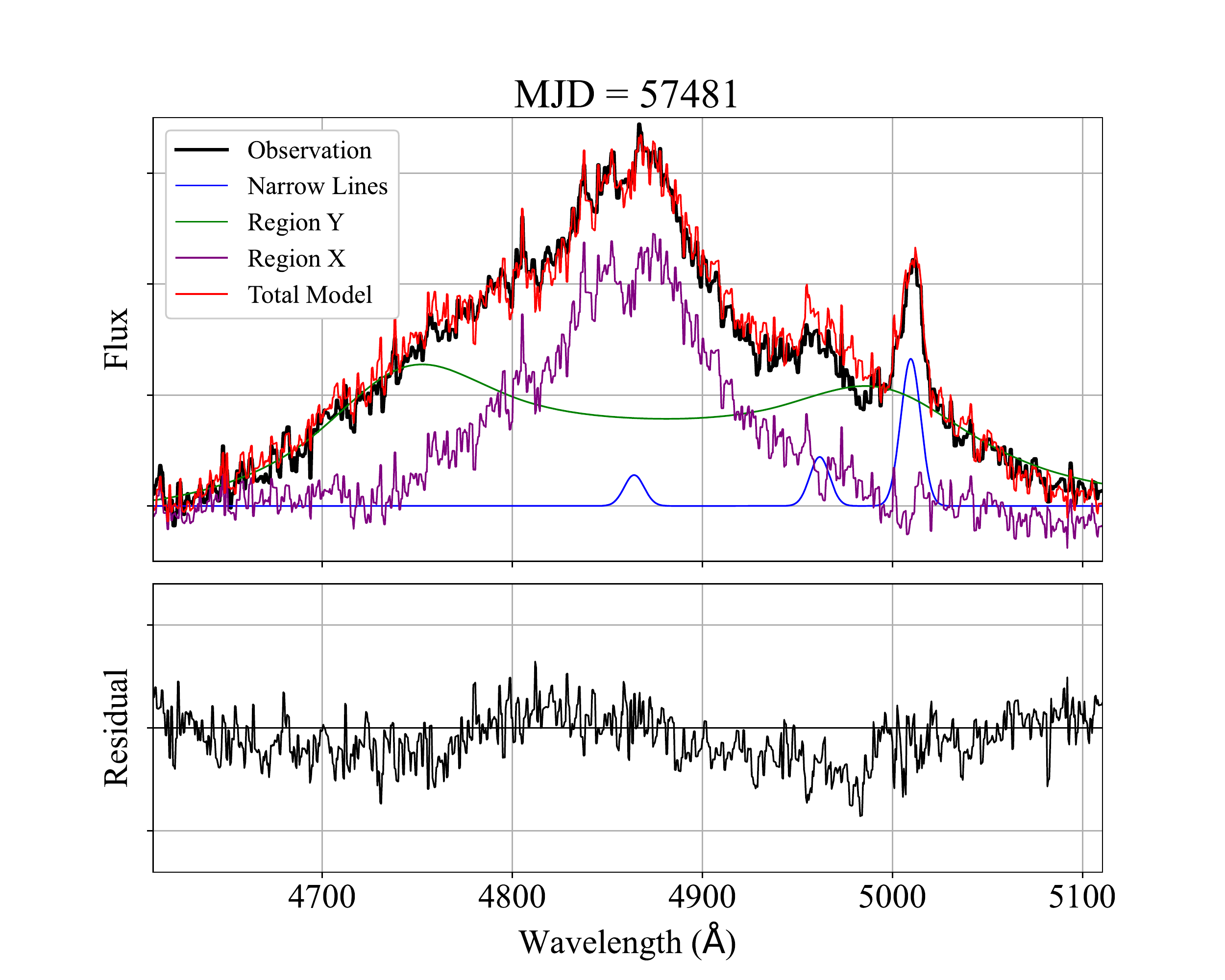}
    \includegraphics[width=0.9\columnwidth]{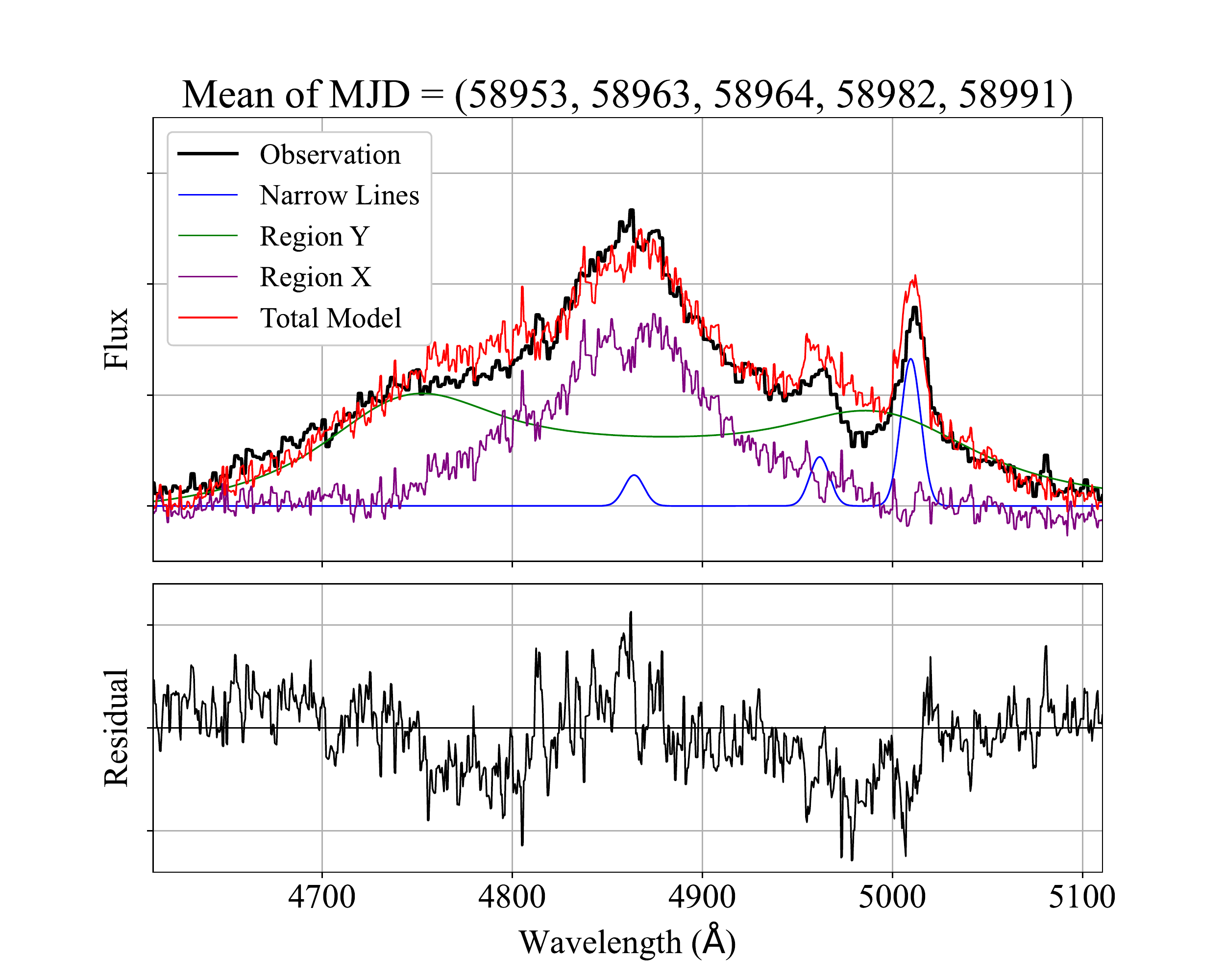}
    \caption{
    Fitting results of the H$\beta$ emission line using two components: the variable component as Region~X and the model of \citet{1989ApJ...339..742C}) as Region~Y. Each panel, from top to bottom, displays the H$\beta$ and [OIII] emission lines with the continuum subtracted for MJD=53848, 57481, and the mean of 58953, 58963, 58964, 58982, and 58991, respectively. In each panel, the upper section illustrates the fitting results of each component, while the lower section presents the fitting residuals. The black line indicates the observed flux, the blue line denotes the narrow component, the green line corresponds to the Region~Y component, the purple line to the Region~X component, and the red line represents the sum of all model components. All panels are displayed with the same grid size.}
    \label{fig:Hbeta_twocomponents_fitting}
\end{figure}

Additionally, utilizing the parameters presented in Figure~\ref{fig:Hbeta_modelfitting}, we verified that by simply optimizing the magnification factors of the components from Region~X (the variable component) and Region~Y (based on the model of \cite{1989ApJ...339..742C}), the H$\beta$ profile of J1258 for all observed periods can be accurately reproduced. Figure~\ref{fig:Hbeta_twocomponents_fitting} displays the fitting results for the dimmest, brightest, and intermediate periods, chosen from spectra with high signal-to-noise ratios. This outcome indicates that our model is effectively able to explain the observed phenomena.

\subsection{Origin of the two components of the BLR}
Based on the structure shown in Figure~\ref{fig:BLR_structure}, we discuss the origins of the gas in the BLR from two major theories: one is that the gas is supplied from disk winds from accretion disks \citep[e.g.,][]{1982ApJ...263...79G, 2000ApJ...530L..65N, 2003ApJ...589L..13N, 2011A&A...525L...8C} and the other is that it is derived from accreted components from the dust torus \citep[e.g.,][]{2009NewAR..53..140G, 2010ASPC..427...68G, 2012MNRAS.420..320H, 2013ApJ...769...30G, 2017NatAs...1..775W}. The disk-wind origin theory is based on a model in which outflow generated from the accretion disk according to an increase of the mass accretion rate winds up the gas, which is ionized by the emission from the inner accretion disk. This model suggests that no BLR is produced below a certain threshold because the disk wind depends on the mass accretion rate \citep{2014MNRAS.438.3340E}. This dependency is sometimes used as the reason for the appearance/disappearance of the BLR with changes in luminosity in the CSQs \citep[e.g.,][]{2015ApJ...800..144L}. On the other hand, the dust torus origin model assumes that BLR is continuously connected with the dust torus, where the ionized component becomes the BLR during mass accretion to the accretion disk.

We suggest that the results of this study can be used to validate the model of the origin of the BLR observationally. First, assuming that disk winds from the accretion disk are the origin of the BLR, strong outflow is expected. Features of outflow are detectable by velocity-resolved RM, which predicts that the red side of the BEL follows the blue side. In other words, the disk wind model is not dominant if the velocity-resolved RM does not detect outflow. The independent variation of the BEL from the continuum variation also excludes the disk wind model, because these two should be correlated in this model. We will now discuss the origin of the Region~X/Y in terms of two aspects: velocity-resolved RM and response to continuum variation, respectively.

\begin{figure}
    \includegraphics[width=\columnwidth]{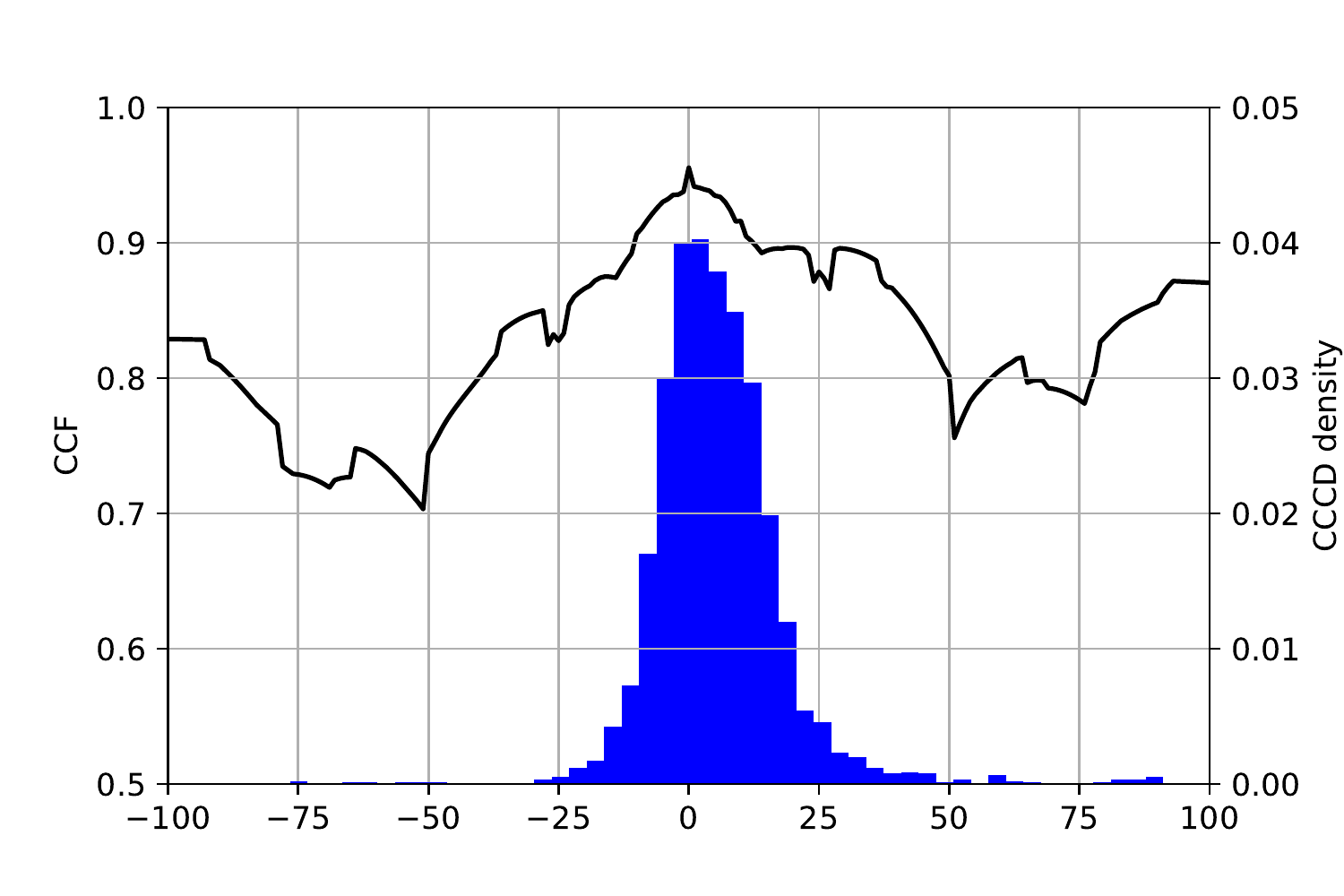}
    \caption{
    ICCF of blue wings relative to red wings of $\rm{H\beta}$, and the distribution of ICCF center.}
    \label{fig:lag_red_blue_ICCF}
\end{figure}

To confirm the origin of Region~X, we conducted a simple velocity-resolved RM. We calculated the cross-correlation function for each of the light curves of the red wing  (4701.3--4861.3 \r{A}) and the blue wing (4861.3--5021.3 \r{A}) components of $\rm{H\beta}$. As a result, the measured time-lag is $4.1^{+3.5}_{-3.1}$ days (Figure~\ref{fig:lag_red_blue_ICCF}). If the gas in Region~X was come from disk winds, the outflow reach Region~X in $0.37$ pc from the center in 20 years, which means the velocity is about $1\%$ of the light speed. When there is strong outflow, the red wing should lag behind the blue wing, because the part far from the observer is moving away \citep{2009NewAR..53..140G}. Here, it is unlikely that gas in Region~X was supplied by outflow, as no significant time-lags could be measured for the red and blue wings. In addition, the evidence of strong inflow was not observed either, suggesting that gas in Region~X doesn't have significant radial motion.

Because no significant radial motion was detected in Region~X, we can consider that the gas existed near the torus before being ionized by the increase in central luminosity \citep[see also][]{2023ApJ...943...63N}. The radius of the ionizing region is thought to be proportional to the central luminosity to the power of 1/2; the BLR's ``breathing'' phenomena have been reported for several objects \citep[e.g.,][]{2020ApJ...903...51W}. Indeed, long-term optical/near-infrared observation reports that the inner radius of the torus of a CSQ, Mrk 590, decreases rapidly following a reduction in the AGN luminosity \citep{2020MNRAS.491.4615K}. The dust sublimation radius measured by the time-lag of $W1$ is about twice the radius of Region~X. The time-lag of $W1$ is mainly dominated by the time-lag when J1258 turns to dim, which may reflect the sublimation radius after it brightens. It is plausible to assume that the dust sublimation radius increased by a factor of 2 during the RM observation period since the continuum luminosity increased 4-5 times during the observation period. Therefore, we conclude that the origin of the gas in Region~X can be said to be the sublimated torus, which was ionized passively with the luminosity variation.

We note, however, that the time lag measurements in this study may underestimate the inner edge radius of the torus. Examining the JAVELIN light curves for the estimated time lag between the continuum and the $W1$ and $W2$ bands (see Figure~\ref{fig:lc_JAVELIN}), the predicted increase/decrease pattern and the actual observation are different, especially in the $W2$ band (the JAVELIN light curve shows a decline around MJD=58500 days, while the actual observation in the $W2$ band continues to increase). This potential underestimation of the time lag may result from the limited observation period. However, even if the actual inner radius of the dust torus measured from the time lag is 2-3 times larger than that in this paper, the inner radius of the dust torus can be expected to have also increased 6-7 times since the object has brightened approximately 40 times over 40 years. Therefore, the gas comprising the BLR observed today is likely the former dust torus.

Region~Y is unlikely to have been formed by outflow, as it is not affected by variation in central luminosity. The asymmetric double-peak component suggests that the gas distribution is biased, formed within a relatively short timescale (< gas circling timescale: $\sim 200$ years). We hypothesize that the dust clumps have been tidally disrupted to create an inhomogeneous gas distribution over a short period \citep{2017NatAs...1..775W}. In other words, Region~Y was formed independently of the state of the accretion disk and thus is less affected by the state transitions.

In summary, the BLR gas in J1258 has two very different positional regions, both of which are likely sourced from the dust torus. Although both originate from the dust torus, we conclude that they have different mechanisms of their origin. One is that a large variation due to state transition changes the radius of being ionized to extend the BLR (Region~X). The other is the ionization of components that enter inside the sublimation radius by accretion of the dust torus (Region~Y).

\section{Conclusions} \label{sec:Conclusion}
In this research, we conducted optical and mid-IR RM and X-ray follow-up observation of an extremely variable CSQ; J1258. This is the first result of RM of the most extremely variable quasar. Our conclusions are as follows.
\begin{itemize}
    \item Black hole mass of J1258 is $10^{9.64^{+0.11}_{-0.20}}\rm{M_\odot}$.
    \item The brightening event can be explained by the state transition of the accretion disk.
    \item The location of the BLR near the central core is inferred from the RM results (Figure~\ref{fig:BLR_structure}). There are two components in the BLR, each located at 0.29 pc and 0.37 pc from the center.
    \item The origin of the ionized gas in Region~X in the BLR is likely to be the sublimated dust torus. The origin of the ionized gas in Region~Y is likely to be accretion from dust torus.
\end{itemize}

\section*{Data Availability}





The data underlying this article will be shared on reasonable request to the corresponding author.

\section*{Acknowledgements}
We express our gratitude to the anonymous reviewer for their insightful comments and valuable suggestions, which have significantly contributed to enhancing the quality of this paper.

This work is financially supported by JSPS KAKENHI grant numbers 22J13428 (S.N.); 22K20391 and 23K13154 (S.Y.). S.Y. is grateful for support from RIKEN Special Postdoctoral Researcher Program. MO was supported by JSPS Grants-in-Aids for Scientific Research(C) (JP19K03914 and 22K03675). This work is partly supported by JSPS KAKENHI Grant Number JP18H05439 and JP20K14521.

We would like to express our gratitude to Dr. Miho Kawabata for support during the observations.

Funding for the Sloan Digital Sky 
Survey IV has been provided by the 
Alfred P. Sloan Foundation, the U.S. 
Department of Energy Office of 
Science, and the Participating 
Institutions. 

SDSS-IV acknowledges support and 
resources from the Center for High 
Performance Computing  at the 
University of Utah. The SDSS 
website is www.sdss4.org.

SDSS-IV is managed by the 
Astrophysical Research Consortium 
for the Participating Institutions 
of the SDSS Collaboration including 
the Brazilian Participation Group, 
the Carnegie Institution for Science, 
Carnegie Mellon University, Center for 
Astrophysics | Harvard \& 
Smithsonian, the Chilean Participation 
Group, the French Participation Group, 
Instituto de Astrof\'isica de 
Canarias, The Johns Hopkins 
University, Kavli Institute for the 
Physics and Mathematics of the 
Universe (IPMU) / University of 
Tokyo, the Korean Participation Group, 
Lawrence Berkeley National Laboratory, 
Leibniz Institut f\"ur Astrophysik 
Potsdam (AIP),  Max-Planck-Institut 
f\"ur Astronomie (MPIA Heidelberg), 
Max-Planck-Institut f\"ur 
Astrophysik (MPA Garching), 
Max-Planck-Institut f\"ur 
Extraterrestrische Physik (MPE), 
National Astronomical Observatories of 
China, New Mexico State University, 
New York University, University of 
Notre Dame, Observat\'ario 
Nacional / MCTI, The Ohio State 
University, Pennsylvania State 
University, Shanghai 
Astronomical Observatory, United 
Kingdom Participation Group, 
Universidad Nacional Aut\'onoma 
de M\'exico, University of Arizona, 
University of Colorado Boulder, 
University of Oxford, University of 
Portsmouth, University of Utah, 
University of Virginia, University 
of Washington, University of 
Wisconsin, Vanderbilt University, 
and Yale University.

Guoshoujing Telescope (the Large Sky Area Multi-Object Fiber Spectroscopic Telescope LAMOST) is a National Major Scientific Project built by the Chinese Academy of Sciences. Funding for the project has been provided by the National Development and Reform Commission. LAMOST is operated and managed by the National Astronomical Observatories, Chinese Academy of Sciences.

This work made use of data supplied by the UK Swift Science Data Centre at the University of Leicester.



The CSS survey is funded by the National Aeronautics and Space
Administration under Grant No. NNG05GF22G issued through the Science
Mission Directorate Near-Earth Objects Observations Program.  The CRTS
survey is supported by the U.S.~National Science Foundation under
grants AST-0909182.






\bibliographystyle{mnras}
\bibliography{main} 




\appendix

\section{}
\begin{figure*}
	\includegraphics[width=2\columnwidth, bb = 0 180 1080 1620]{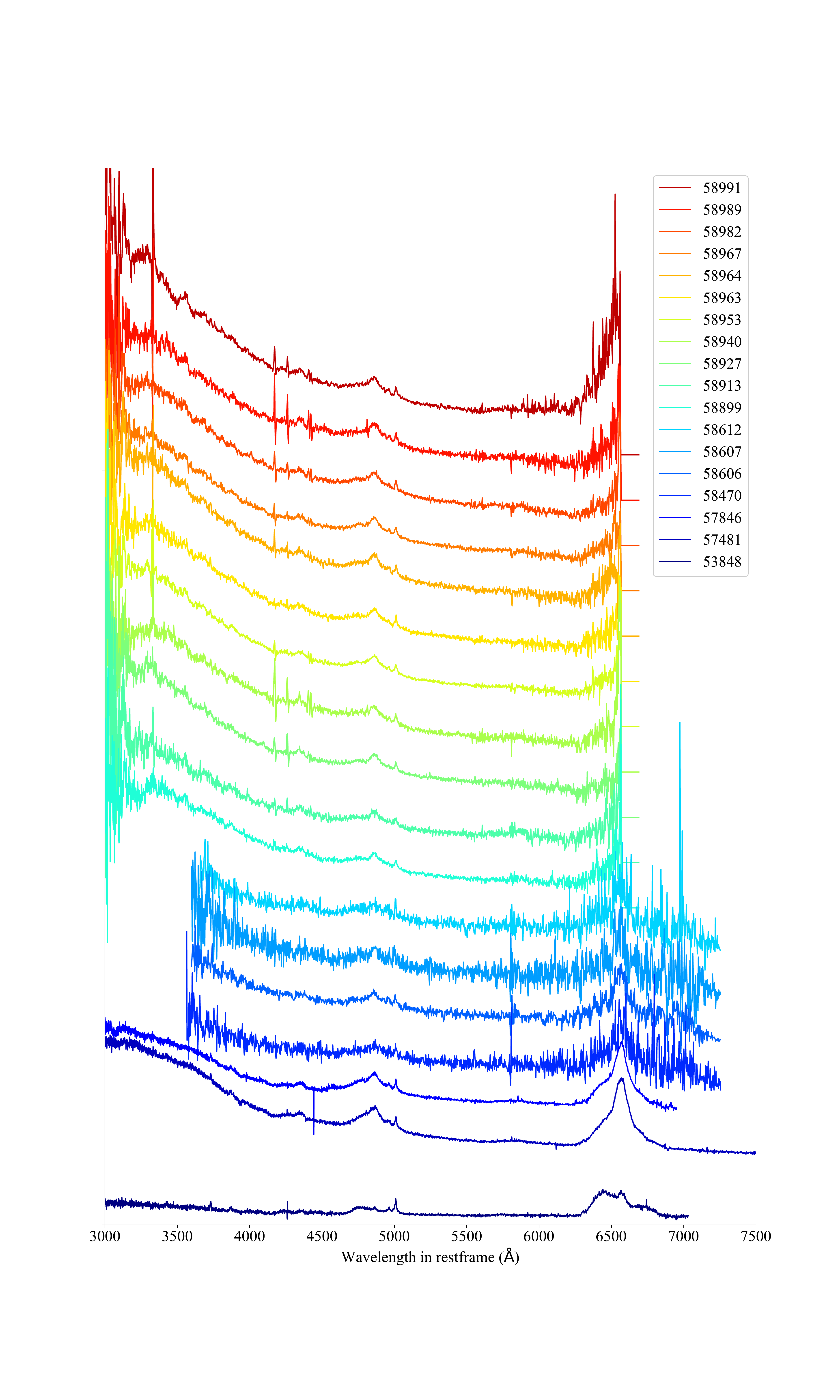}
    \caption{Spectra used in this study. For better visibility, constants have been added to the baseline for each observation. The colours correspond to the date of observation.}
    \label{fig:all_spectra}
\end{figure*}

\begin{figure*}
	\includegraphics[width=1.3\columnwidth]{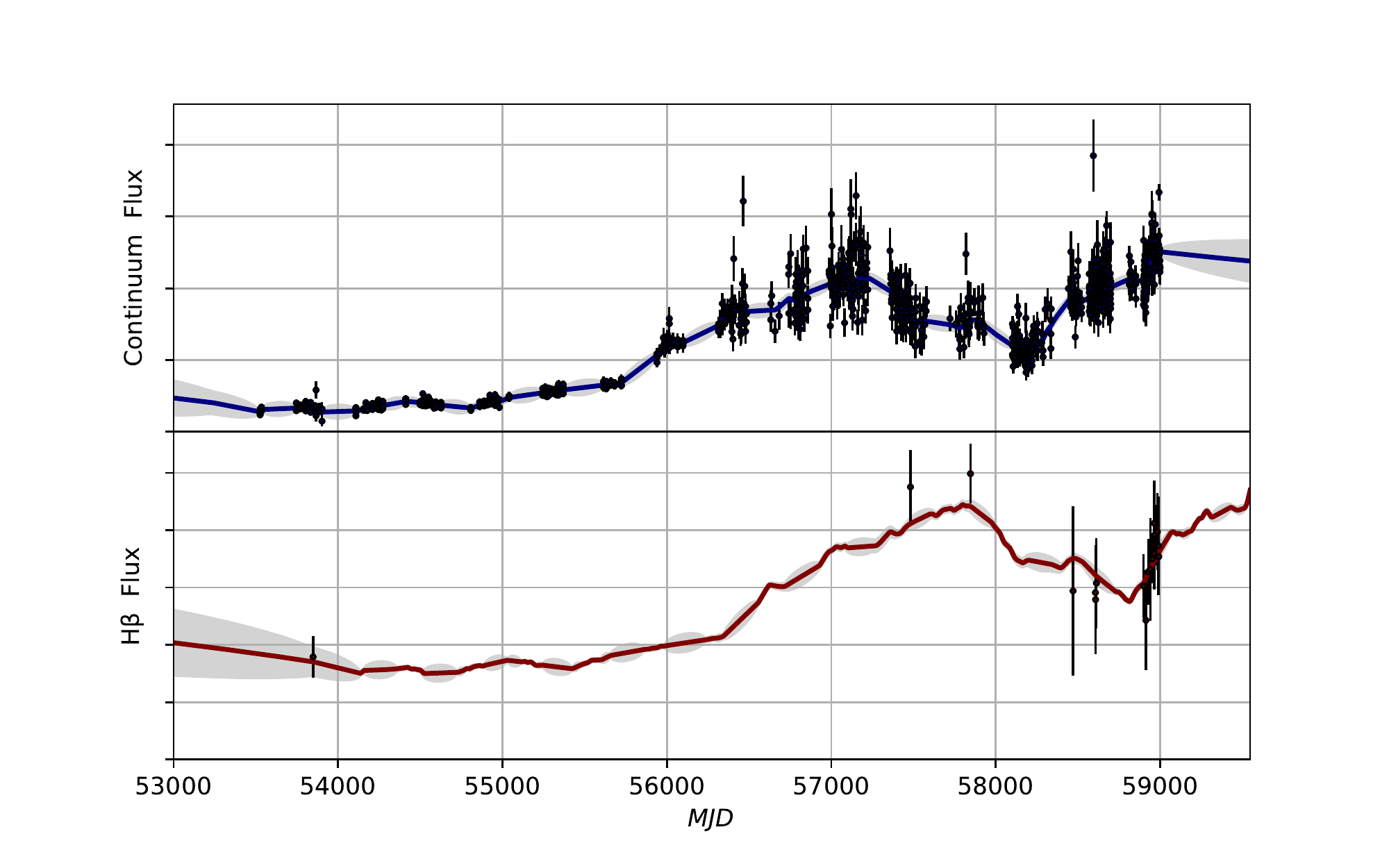}
	\includegraphics[width=1.3\columnwidth]{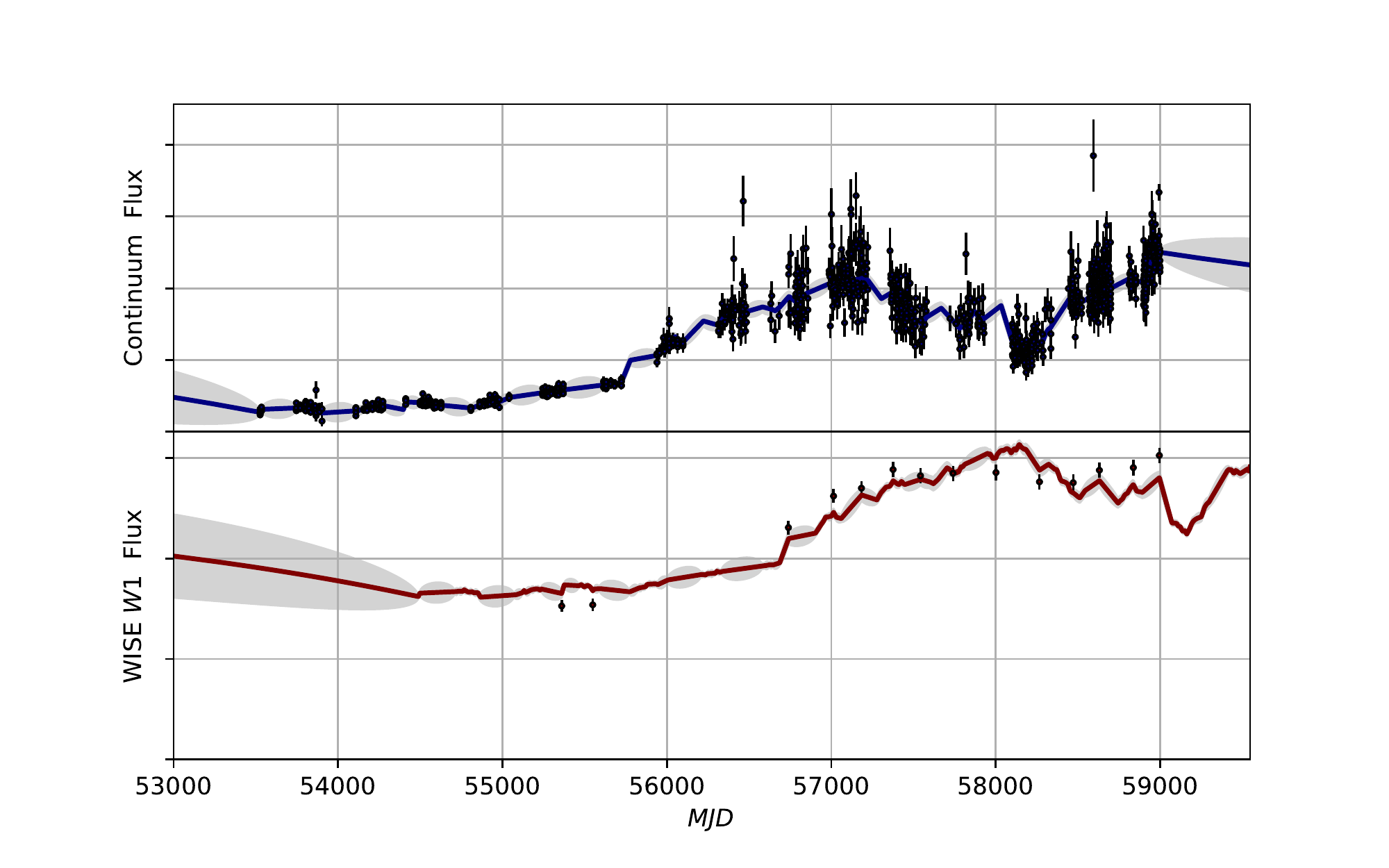}
	\includegraphics[width=1.3\columnwidth]{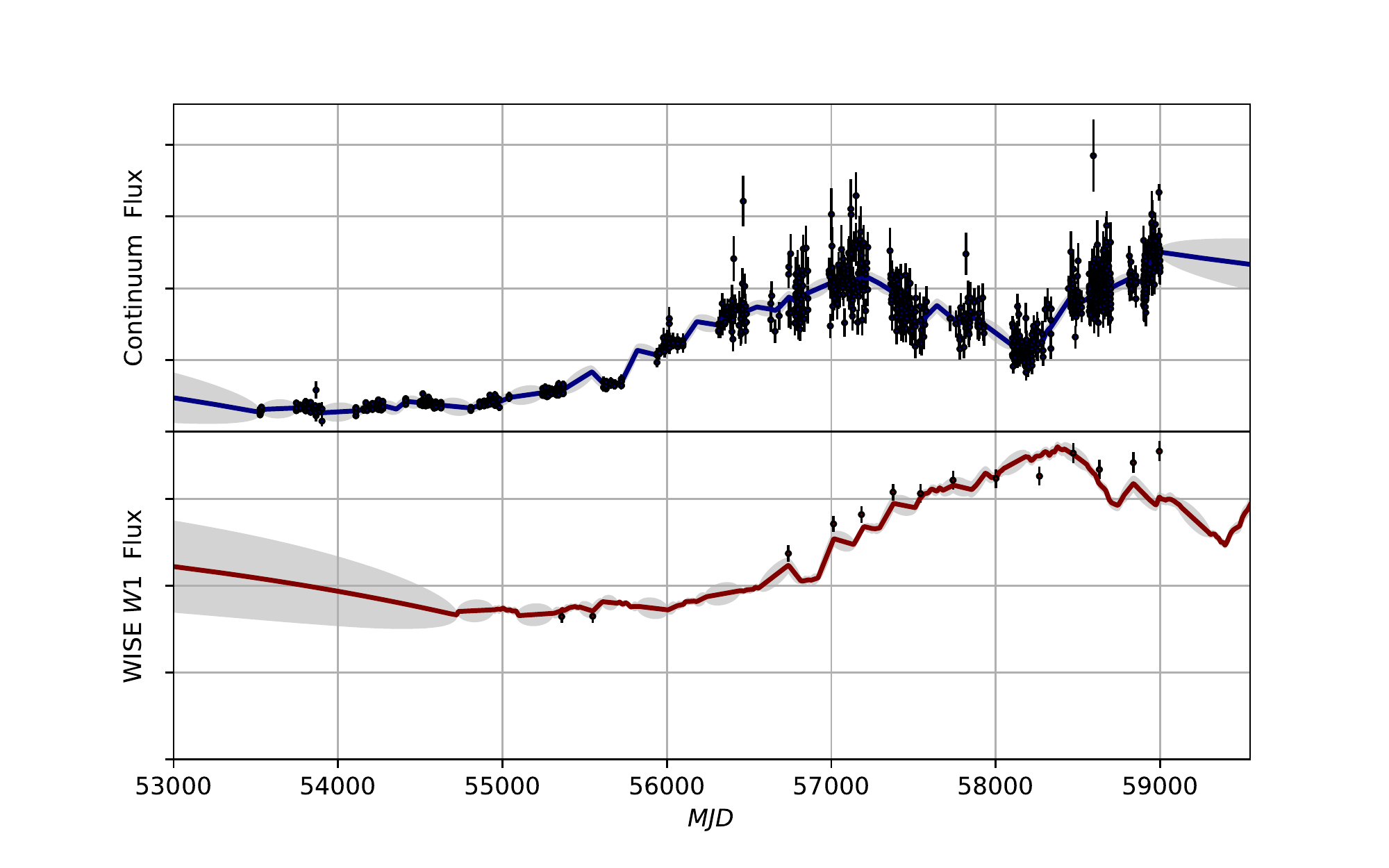}

    \caption{The light curves of continuum (top), $\rm{H\beta}$ (the second), WISE $W1$ (the third), and WISE $W2$ (bottom) of J1258.}
    \label{fig:lc_JAVELIN}
\end{figure*}

In Figure~\ref{fig:all_spectra}, we show all spectra used in this study. In Table~\ref{tab:obslog_table}, we summarize the information of each spectrum and the result of the spectral fitting. 
In Figure~\ref{fig:lc_JAVELIN}, we show the estimated light curves produced by JAVELIN.

\begin{table*}
	\centering
	\caption{List of the spectroscopic observations we use here.}
	\label{tab:obslog_table}
	\begin{tabular}{lcccccccr} 
		\hline
		Observation Date & Instrument & Exposure Time & Standard Star & $\lambda L_{5100 \text{\AA}}$ & $\rm{H\beta}$ \\
		MJD (MM-DD-YYYY) & & second $\times$ number && $10^{44}$ erg/s & $10^{42}$ $\rm{erg/s}$ \\
		\hline
	53848 (04-23-2006) &	SDSS        & $3000 \times 3$  & -          & $ 2.41 \pm 0.02 $ & $ 7.4 \pm 1.5 $ \\
        57481 (04-03-2016) &	SDSS        & $3600 \times 4$  & -          & $ 10.37 \pm 0.02 $ & $ 19.7 \pm 2.7 $ \\
        57846 (04-03-2017) &	LAMOST      & $5400 \times 3$  & -          & $ 8.69 \pm 0.02 $ & $ 20.6 \pm 2.2 $ \\
        58470 (12-18-2018) &	MALLS       & $1200 \times 5$  & HZ44       & $ 6.67 \pm 0.09 $ & $ 11.6 \pm 6.1 $ \\
        58606 (05-03-2019) &	MALLS       & $1200 \times 5$  & BD+33d2642 & $ 7.63 \pm 0.03 $ & $ 12.3 \pm 2.6 $ \\
        58607 (05-04-2019) &	MALLS       & $1200 \times 4$  & BD+33d2642 & $ 7.22 \pm 0.05 $ & $ 12.1 \pm 3.9 $ \\
        58612 (05-09-2019) &	MALLS       & $1200 \times 4$  & HZ44       & $ 7.44 \pm 0.05 $ & $ 12.4 \pm 3.3 $ \\
        58899 (02-20-2020) &	KOOLS-IFU   & $600 \times 10$  & HZ44       & $ 8.06 \pm 0.03 $ & $ 12.6 \pm 2.3 $ \\
        58913 (03-05-2020) &	KOOLS-IFU   & $600 \times 6 $  & HZ44       & $ 7.66 \pm 0.05 $ & $ 10.2 \pm 3.6 $ \\
        58927 (03-19-2020) &	KOOLS-IFU   & $600 \times 8 $  & HZ44       & $ 9.60 \pm 0.03 $ & $ 13.4 \pm 2.4 $ \\
        58940 (04-01-2020) &	KOOLS-IFU   & $600 \times 8 $  & HZ44       & $ 9.87 \pm 0.03 $ & $ 13.2 \pm 3.7 $ \\
        58953 (04-14-2020) &	KOOLS-IFU   & $600 \times 8 $  & HZ44       & $ 11.05 \pm 0.03 $ & $ 14.6 \pm 1.7 $ \\
        58963 (04-24-2020) &	KOOLS-IFU   & $600 \times 7 $  & HZ44       & $ 11.29 \pm 0.03 $ & $ 14.9 \pm 2.6 $ \\
        58964 (04-25-2020) &	KOOLS-IFU   & $600 \times 6 $  & HZ44       & $ 12.70 \pm 0.03 $ & $ 16.4 \pm 3.1 $ \\
        58967 (04-28-2020) &	KOOLS-IFU   & $600 \times 8 $  & HZ44       & $ 11.30 \pm 0.03 $ & $ 16.0 \pm 2.2 $ \\
        58982 (05-13-2020) &	KOOLS-IFU   & $600 \times 8 $  & HZ44       & $ 11.21 \pm 0.03 $ & $ 16.6 \pm 2.8 $ \\
        58989 (05-20-2020) &	KOOLS-IFU   & $600 \times 4 $  & HZ44       & $ 11.77 \pm 0.05 $ & $ 15.2 \pm 3.5 $ \\
        58991 (05-22-2020) &	KOOLS-IFU   & $600 \times 7 $  & HZ44       & $ 12.09 \pm 0.03 $ & $ 14.5 \pm 2.5 $ \\
	\hline
	\end{tabular}
\end{table*}

\begin{figure*}
	\includegraphics[height=0.9\textheight]{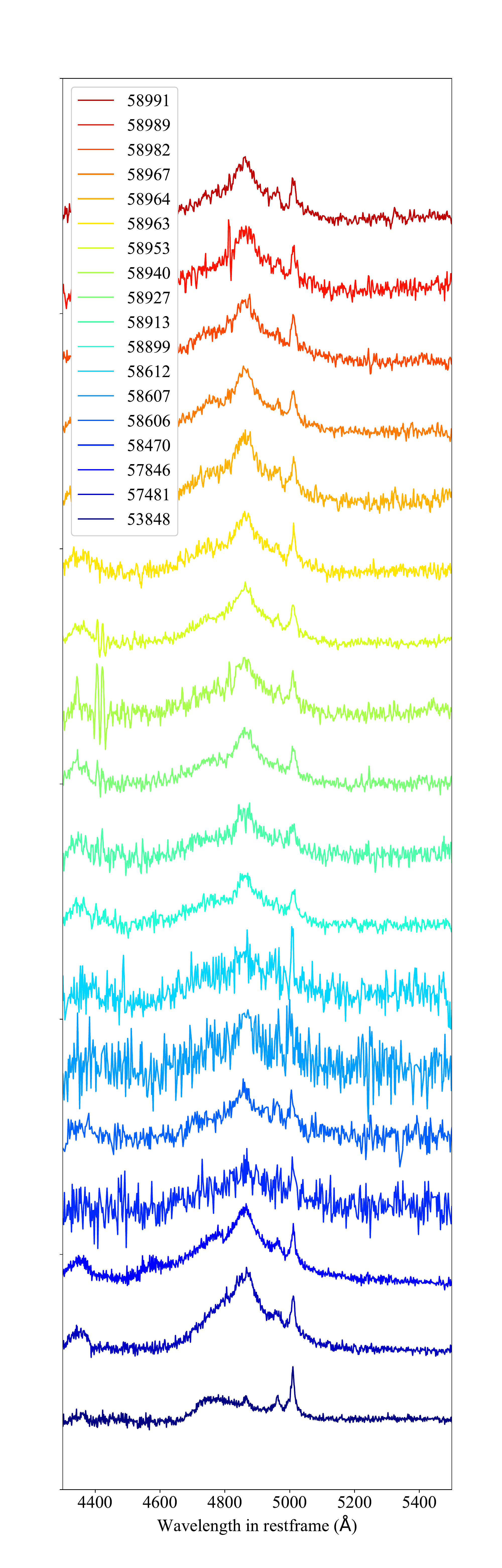}
    \caption{Spectra around the $\rm H\beta$, with the continuum of each spectrum subtracted. For better visibility, constants have been added to the baseline for each observation. The colours correspond to the date of observation.}
    \label{fig:all_Hbeta}
\end{figure*}

\bsp	
\label{lastpage}
\end{document}